\documentclass[onecolumn]{emulateapj}
\usepackage{graphicx}
\usepackage{amsfonts,amsmath,amssymb,mathrsfs}
\usepackage{color}

\newcommand{\be}{\begin{eqnarray}}
\newcommand{\ee}{\end{eqnarray}}

\begin{document}

\title{Testing the nature of the supermassive black hole candidate in S\lowercase{gr}A$^*$\\ 
with light curves and images of hot spots}

\author{Zilong Li, Lingyao Kong, and Cosimo Bambi\footnote{Corresponding author: bambi@fudan.edu.cn}}

\affil{Center for Field Theory and Particle Physics \& Department of Physics, Fudan University, 200433 Shanghai, China}

\begin{abstract}
General relativity makes clear predictions about the spacetime geometry around
black holes. In the near future, new facilities will have the capability to explore
the metric around SgrA$^*$, the supermassive black hole candidate at the Center 
of our Galaxy, and open a new window to test the Kerr black hole hypothesis. In 
this paper, we compute light curves and images associated with compact emission 
regions (hot spots) orbiting around Kerr and non-Kerr black holes. We study how 
the analysis of the properties of the radiation emitted by a hot spot can be used to 
test the Kerr nature of SgrA$^*$. We find that the sole observation of the hot spot 
light curve can at most constrain a combination of the black hole spin and of possible 
deviations from the Kerr solution. This happens because the same orbital frequency 
around a Kerr black hole can be found for a non-Kerr object with a different spin 
parameter. Second order corrections in the light curve due to the background geometry 
are typically too small to be identified. 
While the observation of the hot spot centroid track can potentially bound
possible deviations from the Kerr solution, that is out of reach for the near future 
VLTI instrument GRAVITY.
The Kerr black hole 
hypothesis could really be tested in the case of the discovery of a radio pulsar in 
a compact orbit around SgrA$^*$. Radio observations of such a pulsar would 
provide precise estimates of the mass and the spin of SgrA$^*$, and the combination 
of these measurements (probing the weak field) with the hot spot light curve 
information (probing the strong field) may constrain/find possible deviations 
from the Kerr solution with quite good precision.
\end{abstract}



\section{Introduction}

In 4-dimensional general relativity, uncharged black holes (BHs) are described
by the Kerr solution and are completely characterized by only two quantities:
the mass $M$ and the spin parameter $a = J/M$, where $J$ is the BH spin angular 
momentum. Astrophysical BH candidates are thought to be the Kerr BHs predicted 
in general relativity simply because they cannot be explained otherwise without 
introducing new physics. Stellar-mass BH candidates in X-ray binary systems 
have a mass of $M \approx 5 - 20$~$M_\odot$, which is too high for a neutron or 
quark star for any plausible matter equation of state~\citep{bh1,bh2}. At least 
some supermassive BH candidates at the center of galaxies are definitively too
massive, compact, and old to be a cluster of compact non-luminous objects, as the 
cluster lifetime due to evaporation and physical collisions would be shorter than 
the age of these systems~\citep{bh3}. The non-observation of electromagnetic 
radiation emitted by the possible surface of these objects may also be interpreted 
as an evidence for the presence of an event horizon~\citep{ah0,ah1,ah2}, but 
only if we exclude the possibility of new physics~\citep{ah3,ah4,ah5}. Despite these
arguments, there is no direct indication that the spacetime geometry around BH 
candidates is described by the Kerr metric and so far the predictions of general 
relativity have been tested only in the weak field regime, while the validity of the 
theory in strong gravitational fields is totally unexplored.

In the last few years, the possibility of testing the Kerr nature of BH candidates by 
studying the properties of the electromagnetic radiation emitted by the gas in the 
accretion disk has been discussed by several authors and it has become a quite 
active line of research [for a review, see e.g.~\citet{rev1,rev2}]. In this context, it is
important to bear in mind two fundamental points, which instead are usually not 
well understood. First, strictly speaking the electromagnetic radiation emitted from 
the accretion disk can only be used to check if the spacetime geometry around 
BH candidates is described by the Kerr metric, not to test the validity of the Einstein 
equations. Indeed, these photons simply follow the null geodesics of the spacetime,
independently of the exact equations determining the background metric. The 
Kerr metric is a solution of the Einstein equations, but it is common to other theories 
of gravity~\citep{v2-kerr1}. If we want to test the validity of the Einstein equations, we 
should study the perturbations around this metric~\citep{v2-kerr2}. Second, contrary 
to common statements in the literature concerning tests of gravity in the strong field 
regime, the observation of features associated to relativistic effects absent in 
Newtonian gravity is not enough to test the Kerr BH hypothesis. The correct approach 
is instead to compare observational data with theoretical predictions in Kerr and 
non-Kerr backgrounds and see if the observations can distinguish the two scenarios. 
That is usually not the case, because there is a degeneracy between the spin and 
possible deviations from the Kerr metric, and a non-Kerr object looks like a Kerr BH
with a different spin.

At present, there are two relatively robust techniques to probe the spacetime 
geometry around BH candidates and constrain possible deviations from the Kerr 
solution; that is, the fit of the thermal spectrum of thin disks~\citep{zhang,cfm1,cfm2} 
and the analysis of the K$\alpha$ iron line profile~\citep{fabian,iron1,iron2}. 
However, it turns out that it is very difficult to test the Kerr BH paradigm, because 
often a non-Kerr BH looks like a Kerr one with a different spin. The disk's thermal 
spectrum and the iron line profile can currently be used to exclude some exotic 
BH alternatives, like some wormholes~\citep{iron3} and some exotic compact 
objects without event horizon~\citep{iron4,joshi}. More realistic scenarios can be 
extremely difficult to test~\citep{cfm-iron}. Some tentative bounds inferred from current 
data are reported, for instance, in~\citet{agn1,cfm3,iron5} and \citet{agn2}.

In addition to the analysis of the disk's thermal spectrum and of the iron line profile,
there are some quite promising techniques for the future, especially if used in 
combinations with other measurements. For instance, the observed
quasi-periodic oscillations (QPOs) in the X-ray flux of stellar-mass BH candidates
may be a very powerful tool to probe the spacetime geometry around these 
objects~\citep{qpo1,qpo2,qpo3}. At present, the exact mechanism responsible
for these phenomena is not understood and different scenarios provide different
results, which makes it not yet possible to use QPOs to test fundamental physics.
The same problem affects the measurements via the estimate of the jet 
power~\citep{nmc1,jet1,jet2,nmc2}: different models lead to different conclusions
and at present we do not know which one is correct. In the future, it will be possible 
to test the nature of astrophysical BH candidates with data not yet available, like
the polarization of X-ray radiation from the accretion disk~\citep{polarization} and
pulsar timing~\citep{v2-pulsar1,v2-pulsar2}.

A promising object to test the Kerr BH paradigm is SgrA$^*$, 
the supermassive BH candidate at the Center of our 
Galaxy. For instance, very long baseline interferometry (VLBI) observations at 
mm wavelengths have recently shown that it will be probably possible to image 
the accretion flow around SgrA$^*$ within about 5 years~\citep{doe1,doe2} and 
open a new window to test general relativity in the strong field regime. 
The main goal of the Event Horizon 
Telescope\footnote{URL: http://www.eventhorizontelescope.org/} is the observation 
of the BH ``shadow'', a dark area over a brighter background~\citep{shk,rohta}. 
While the intensity map of the image of the accretion flow depends also on 
complicated astrophysical processes, the exact shape of the shadow is only 
determined by the background geometry. Indeed, the boundary of the shadow 
corresponds to the apparent photon capture sphere of the background metric as 
seen by a distant observer. Starting from \citet{shadow1} and \citet{bft}, the shape 
of the shadow has been extensively studied to test the metric around SgrA$^*$ 
with future VLBI observations~\citep{shadow2,shadow3,shadow-jp,shadow4,
shadow5,shadow6,shadow7,shadow8,shadow10,shadow9,v2-naoki}. For a 
review on the subject, see e.g. \citet{shrev}.

While the detection of the BH shadow is surely an extremely intriguing goal, 
tests of general relativity will probably require quite high resolutions, not available 
soon [but see \citet{shadow6}]. In the near future, a more promising possibility
could be the observation of compact emission regions (hot spots) orbiting near 
the innermost stable circular orbit (ISCO) of SgrA$^*$. Indeed, the latter exhibits
powerful flares in the X-ray, NIR, and sub-mm bands~\citep{v2-flares,var1,var2,
v2-ff3,v2-flares2,v2-ff2,v2-ff1}; for a review, see e.g. \citet{v2-review}. During the
flares, the flux increases up to a factor 10. A flare typically lasts 1-3 hours and
the rate is of a few events per day. 
Flares seem to show a quasi-periodic substructure with a time scale of about 
20 minutes, but the presence of this substructure is still the subject of intense 
debate and some authors argue that it is not present in the data~\citep{v3-do}.
Several mechanisms have been proposed
to explain these flares, such as the heating of electrons in a jet~\citep{v2-model2},
the adiabatic expansion of a blob of plasma~\citep{v2-model1}, Rossby wave 
instability in the disk~\citep{v2-model3}, and blobs of plasma orbiting the ISCO 
of SgrA$^*$~\citep{v2-model4}. 
In this paper, we will assume that the last scenario of a hot spot near the 
ISCO is the correct one, but at present there is no reason to favor it. The hot
spot model will be tested by
the GRAVITY instrument for the ESO Very Large Telescope Interferometer
(VLTI)~\citep{v2-gravity,v2-gravity-b,v2-gravity2}\footnote{URL: 
http://www.mpe.mpg.de/ir/gravity}.

Computations of images and light curves of hot spots in the Kerr metric for different 
values of the BH spin parameter, orbital radius, and viewing angle were previously
reported in~\citet{sch1}, \citet{sch2,sch3}, \citet{brod1,brod2}, and \citet{v2-model4}, 
where it has been shown that the mass and the spin of the BH may be extracted 
from their signatures in the flux and in the polarization of the hot spot.
The aim of the present paper is to extend the study of a hot spot in a Kerr
background to test the spacetime geometry around SgrA$^*$. We compute images 
and light curves of an optically thick emitting disk orbiting around Kerr and 
non-Kerr BHs for different values of the model parameters. In particular, we want 
to see if it is possible to estimate the value of the spin and constrain deviations 
from the Kerr metric at the same time. As in the case of other approaches, it turns
out that this is quite a difficult job. If we assume that the BH is of the Kerr type, 
the only parameter of the background geometry affecting the strong field regime 
is the spin (the mass can be inferred by dynamical methods, by studying the 
orbital motion of individual stars in the Newtonian regime). Any observable
quantity that is a monotonic function of the spin can be used to infer the latter. If
we relax the Kerr BH hypothesis, the properties of the spacetime metric close to
the compact object depend on the spin as well as on possible deformations from
the Kerr geometry. If the hot spot is orbiting at the ISCO, its light curve essentially 
tells us the ISCO frequency. While in the Kerr metric there is a one-to-one 
correspondence between BH spin and ISCO frequency (if the BH mass is 
known), that is not true any more in a non-Kerr background, because the ISCO 
frequency now depends also on the deformation parameter. The latter introduces 
also some small corrections to the shape of the light curve, but these corrections 
are very small and they can unlikely be used to estimate the BH spin and the
deformation parameter at the same time. The sole observation of the light curve of 
a hot spot orbiting at the ISCO can only select the spacetimes with the same ISCO
frequency. 
The observation of the centroid track may distinguish spacetimes with the 
same ISCO frequency and thus constrain possible deviations from the Kerr 
solution, but a 10~$\mu$as precision like GRAVITY is not enough to do it.
We find that an interesting possibility is the
combination of the light curve information with the possible future observation of 
a pulsar orbiting SgrA$^*$ with a period of a few months. Such a pulsar would allow 
very precise measurements of the BH mass and spin (independently of the nature
of the BH candidate, because it is relatively far from the compact object)~\citep{v2-pulsar2}. 
Once the mass and the spin are known, the ISCO frequency depends only on possible
deviations from the Kerr solution. While the pulsar data might also independently
constrain the BH quadrupole moment, a hot spot would provide more stringent constraints 
on the Kerr geometry, because it does not require that the BH spin is high and 
it is sensitive even to deviations of higher order.

The content of the paper is as follows. In Section~\ref{s-k}, we review the calculations 
of images and light curves of hot spots in a Kerr background. In Section~\ref{s-nk}, 
we apply this approach to the case of non-Kerr spacetimes. In Section~\ref{s-d}, we 
compare the observational properties of hot spots orbiting around Kerr and non-Kerr 
BHs to figure out how future observations can test the Kerr metric around SgrA$^*$. 
Summary and conclusions are reported in Section~\ref{s-c}. Throughout the paper, 
we use units in which $G_{\rm N} = c = 1$.

\vspace{1cm}

\section{Hot spots orbiting Kerr black holes}
\label{s-k}

General relativistic magneto-hydrodynamic simulations of accretion flows onto 
BHs indicate that temporary clumps of matter may be common in the region near 
the ISCO~\citep{v2-grmhd1,v2-grmhd2}. In the case of SgrA$^*$, such a
possibility seems to be supported by the observation of flaring activity in the
X-ray, NIR, and sub-mm bands with a timescale of order the orbital frequency 
at the ISCO radius~\citep{v2-flares,var1,var2,v2-ff3,v2-flares2,v2-ff2,v2-ff1}. 
A similar evidence comes from the X-ray spectrum of stellar-mass BH candidates 
in binary systems, where the observed QPOs in the X-ray flux have a frequency 
comparable to the expected fundamental orbital frequencies of a test-particle orbiting 
near the ISCO radius~\citep{sv1,sv2}. All these arguments support the hot spot model, 
which has been already extensively discussed in the literature in the case of Kerr 
spacetime~\citep{sch1,sch2,sch3,brod1,brod2}. For a 4 million Solar mass Kerr BH, 
the ISCO period ranges from about 30~minutes ($a/M = 0$) to 4~minutes ($a/M = 1$ 
and corotating orbit). In the case of SgrA$^*$, the observed period of the flare 
quasi-periodic substructure ranges from 13 to about 30~minutes.
NIR flares have been reported up to about 45~minutes [see Fig.~12 in~\citet{v2-model4}].
This suggests that the orbital radius of the hot spot is not necessarily always at the 
ISCO, but it can vary and be at larger radii~\citep{v2-trippe}. 
The shorter period ever measured is 
$13\pm2$~minutes, and it may be assumed either the ISCO period or, more 
conservatively, an upper bound for it. In the latter case, with a mass $M = 3.6 \cdot 
10^6$~$M_\odot$ for SgrA$^*$, one obtains $a/M \ge 0.70\pm
0.11$~\citep{v2-trippe}\footnote{In~\citet{v2-5min1} and \citet{v2-5min2}, the 
authors claimed the presence of a quasi-periodic substructure with a period of 
5~minutes, which was interpreted as an indication that SgrA$^*$ is rotating very 
fast, with $a/M$ close to 1. However, the analysis of the same data sets 
in~\citet{v2-no5min} did not find such a short period substructure.}.

The simplest hot spot model, which will be employed throughout this paper, is a 
single region of isotropic and monochromatic emission following a geodesic 
trajectory. Located on the equatorial plane, this hot spot is modeled as an optically 
thick emitting disk of finite radius. The local specific intensity of the radiation is 
chosen to have a Gaussian distribution in the local Cartesian space~\citep{sch3} 
\be
I_{\rm em}(\nu_{\rm em},x) \sim \delta(\nu_{\rm em} - \nu_\star)
\exp\bigg[-\frac{|\textbf{\~{x}}-\textbf{\~{x}}_{\rm spot}(t)|^2}
{2R^2_{\rm spot}}\bigg] \, ,
\label{emissivity}
\ee
where $\nu_{\rm em}$ is the photon frequency measured in the rest-frame of the 
emitter, while $\nu_\star$ is the emission frequency of this monochromatic source. The 
spatial position 3-vector $\textbf{\~{x}}$ is given in pseudo-Cartesian coordinates.
Outside a distance of $4R_{\rm spot}$ from the guiding geodesic trajectory 
$\textbf{\~{x}}_{\rm spot}$, there is no emission. Plausible values are $R_{\rm spot} = 
0.1 - 1.0 M$, but it turns out that the light curves are not very sensitive to the exact spot size 
(see below and Fig.~\ref{fig4}). More precisely, the light curve of the primary image 
is quite independent of it, while the effect of the spot size is more pronounced in the 
light curve of the secondary image, at least when the spot is orbiting very close to 
the compact object. Since we assume all points in the hot spot have the same 4-velocity 
as the geodesic guiding trajectory, one must be careful not to use a too large spot, 
or the point of emission $\textbf{\~{x}}$ can be spatially far enough away from the 
center to make the results unphysical.

The calculation of the electromagnetic emission of a hot spot is, in many aspects,
similar to the computation of the radiation emitted by a thin accretion disk~\citep{li05}. 
The main difference is that the spectrum of the moving spot is time-dependent. First,
we compute the trajectories of the photons backwards in time from the image plane
of the distant observer to the orbital plane of the hot spot. The observer's sky is 
divided into a number of small elements and the ray-tracing procedure provides 
the observed time-dependent flux density from each element. In the special case 
of the Kerr background, one can exploit the properties of the Kerr solution and solve a 
simplified set of differential equations. Since in the next section we will consider 
non-Kerr spacetimes without these properties, we use the code described in \citet{cfm2}, 
which solves the second-order photon geodesic equations by using the fourth-order
Runge-Kutta-Nystr\"{o}m method~\citep{lund}. The initial conditions 
$(t_0, r_0, \theta_0, \phi_0)$ for the photon with Cartesian coordinates $(X,Y)$  
on the image plane of the distant observer and detection time $t_{\rm obs}$ are 
given by~\citep{shadow-jp}
\begin{align}
t_0 &= t_{\rm obs} \, , \label{t0} \\
r_0 &= \sqrt{X^2+Y^2+D^2} \, , \\
\theta_0 &= \arccos\frac{Y\sin i + D\cos i}{r_0} \, , \\
\phi_0 &= \arctan\frac{X}{D\sin i - Y\cos i} \, .
\end{align}
As the initial 3-momentum $\bf{k_0}$ must be perpendicular to the plane of the 
image of the observer, the initial conditions for the 4-momentum of the photon are
\begin{align}
k^r_0 &= -\frac{D}{r_0}|\bf{k_0}| \, ,  \\
k^\theta_0 &= \frac{\cos i -(Y\sin i + D\sin i)
\frac{D}{r_0^2}}{\sqrt{X^2+(D\sin i- Y\cos i)^2}}|\bf{k_0}| \, , \\
k^\phi_0 &= \frac{X\sin i}{X^2+(D\sin i- Y\cos i)^2}|\bf{k_0}| \, , \\
k^t_0 &= \sqrt{(k^r_0)^2+r_0^2(k^\theta_0)^2+r_0^2
\sin^2 \theta_0(k^\phi_0)^2} \, ,
\end{align}
where $D$ is the distance of the observer from the BH and $i$ is the observer 
line of sign with respect to the BH spin. In the numerical calculations, the observer 
is located at the distance $D = 10^6 M$ (where $M$ is the mass of the central 
compact object), which is far enough to assume that the background geometry 
is flat and therefore $k^t_0$ can be inferred from the condition $g_{\mu\nu}
k^\mu k^\nu = 0$ with the metric tensor of a flat spacetime.

\begin{figure}
\begin{center}
\includegraphics[width=8cm]{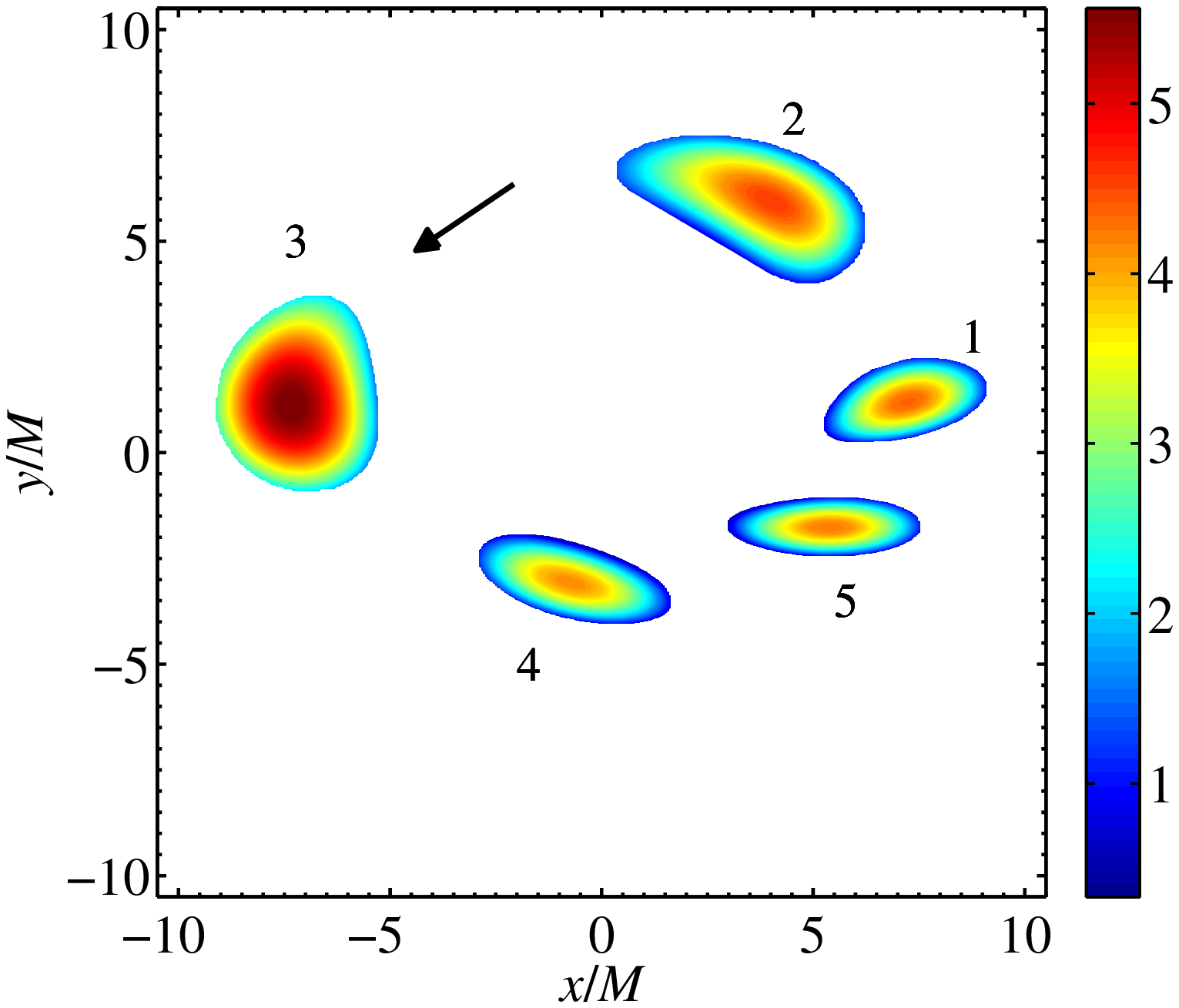}
\includegraphics[width=8cm]{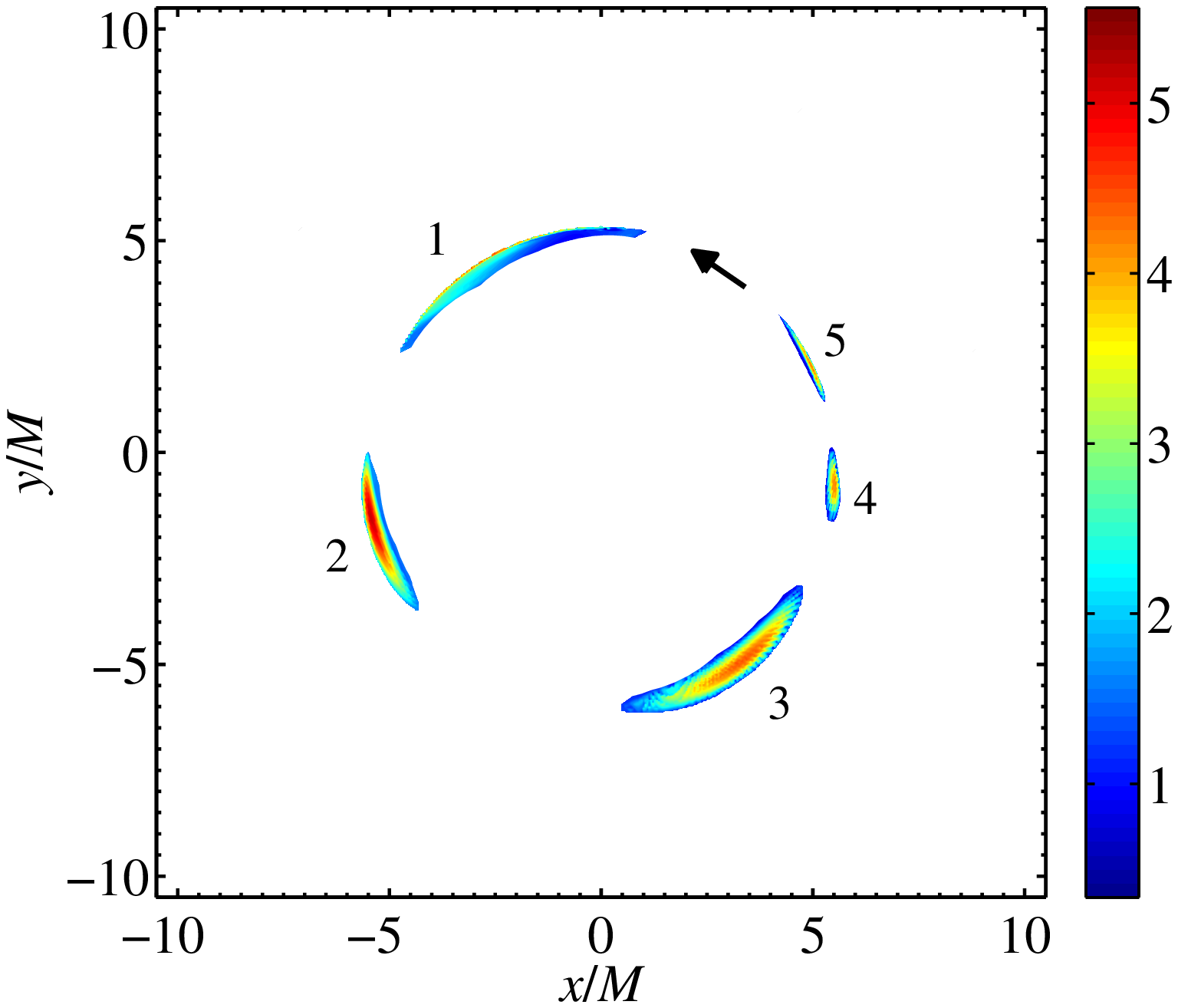}
\end{center}
\caption{Left panel: specific intensity of primary images of a hot spot with a radius of 
$R_{\rm spot} = 0.5 \, M$ orbiting a Schwarzschild BH at the ISCO. The observer's
viewing angle is $i = 60^\circ$ and the time interval between two adjacent spot 
images is $T/5$, where $T$ is the orbital period of the spot. Right panel: as in 
the left panel for the corresponding secondary images. Specific intensity in 
arbitrary units. See the text for more details.}
\label{fig1}
\begin{center}
\includegraphics[width=8cm]{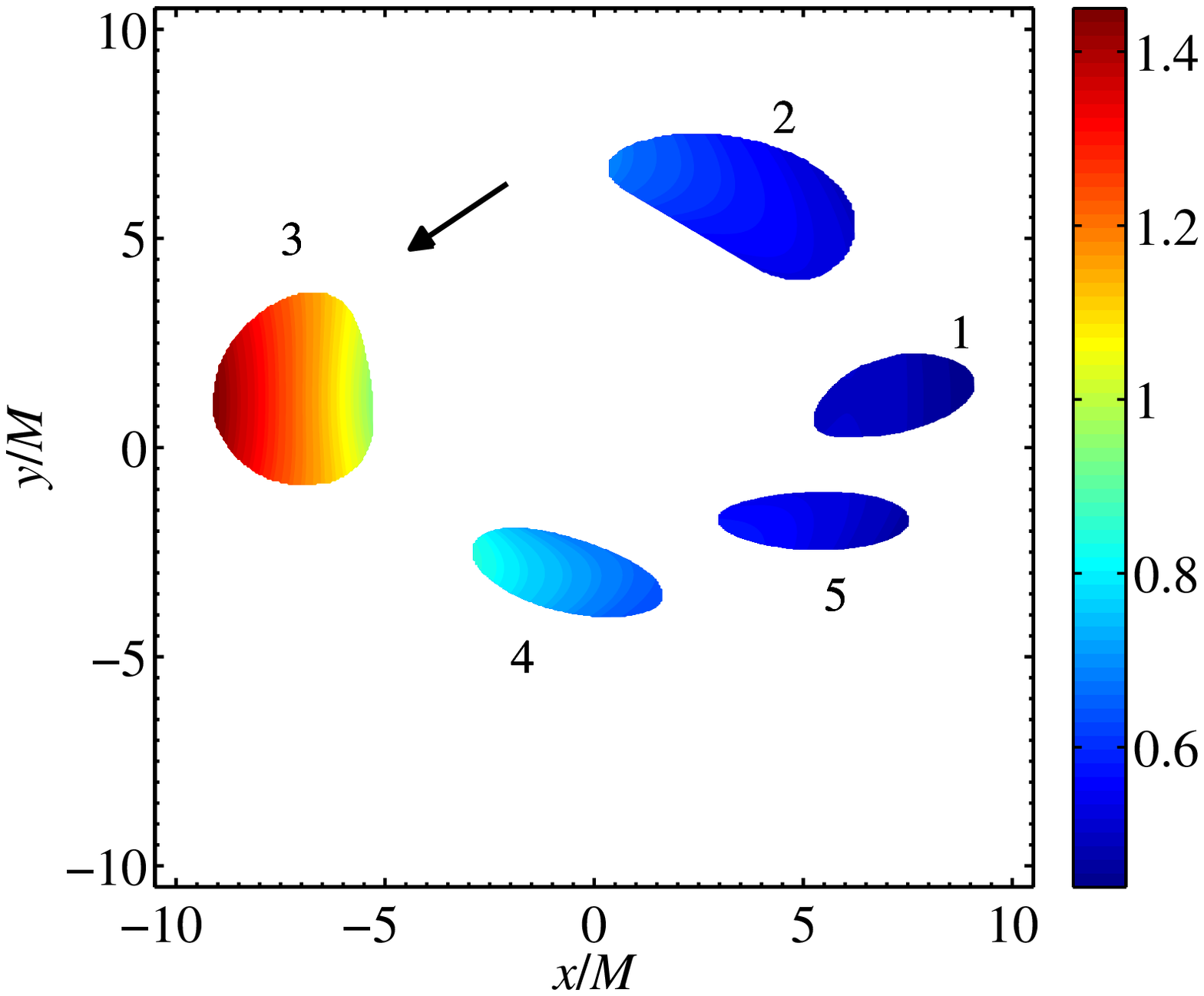}
\includegraphics[width=8cm]{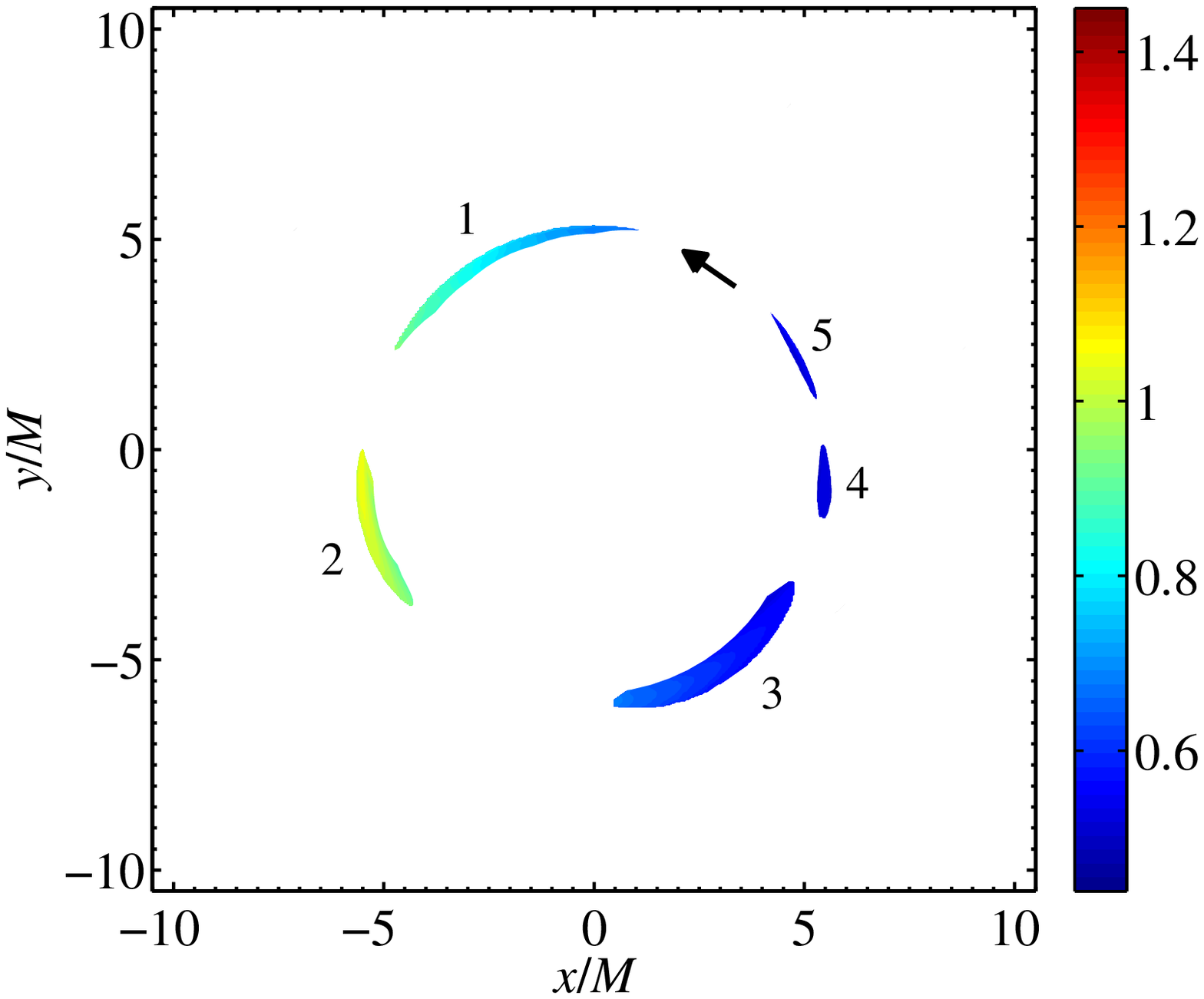}
\end{center}
\caption{Left panel: redshift factor $g$ of primary images of a hot spot with a radius 
of $R_{\rm spot} = 0.5 \, M$ orbiting a Schwarzschild BH at the ISCO. The observer's
viewing angle is $i = 60^\circ$ and the time interval between two adjacent spot 
images is $T/5$, where $T$ is the orbital period of the spot. Right panel: as in 
the left panel for the corresponding secondary images. See the text for more 
details.}
\label{fig1bis}
\begin{center}
\includegraphics[height=7cm]{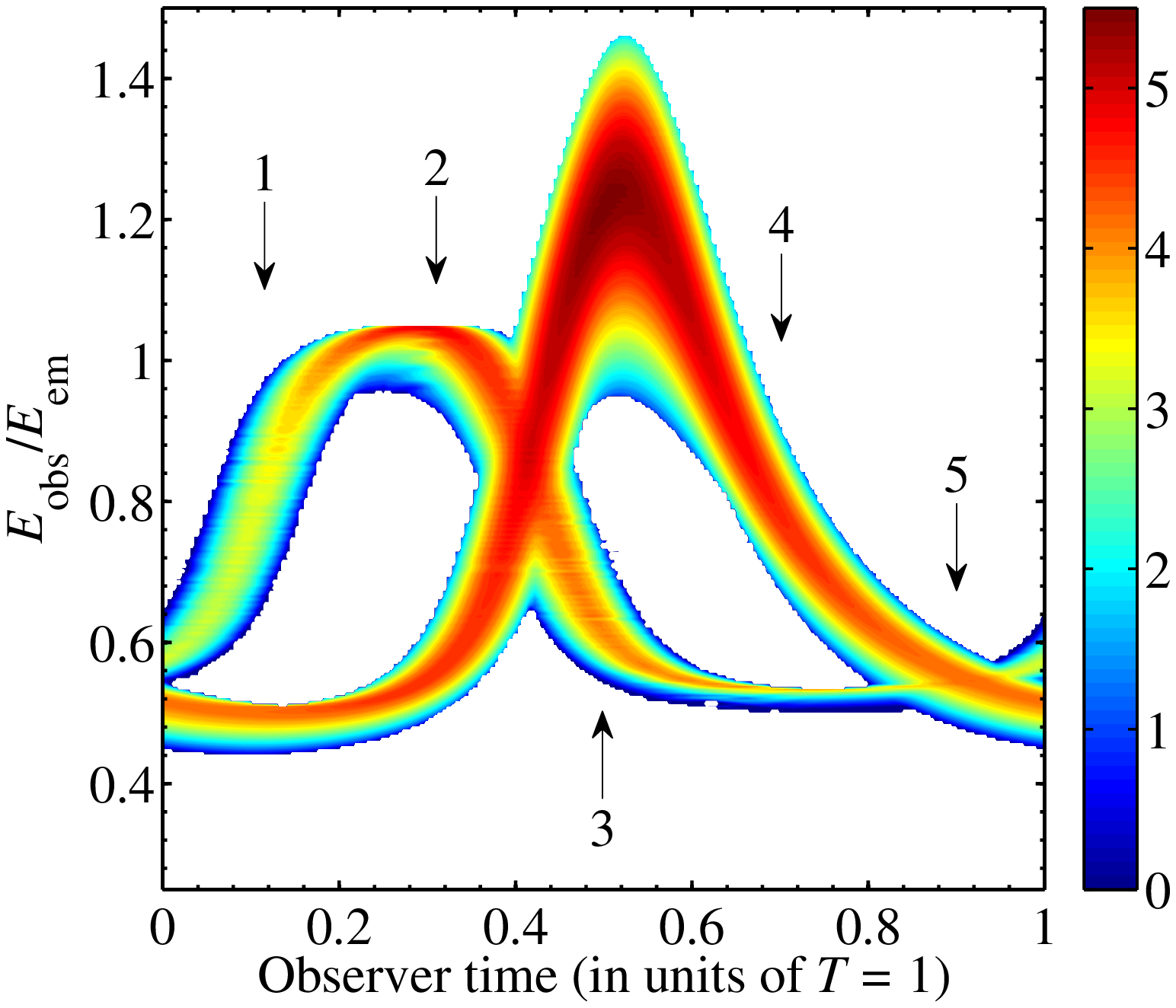}
\includegraphics[height=7cm]{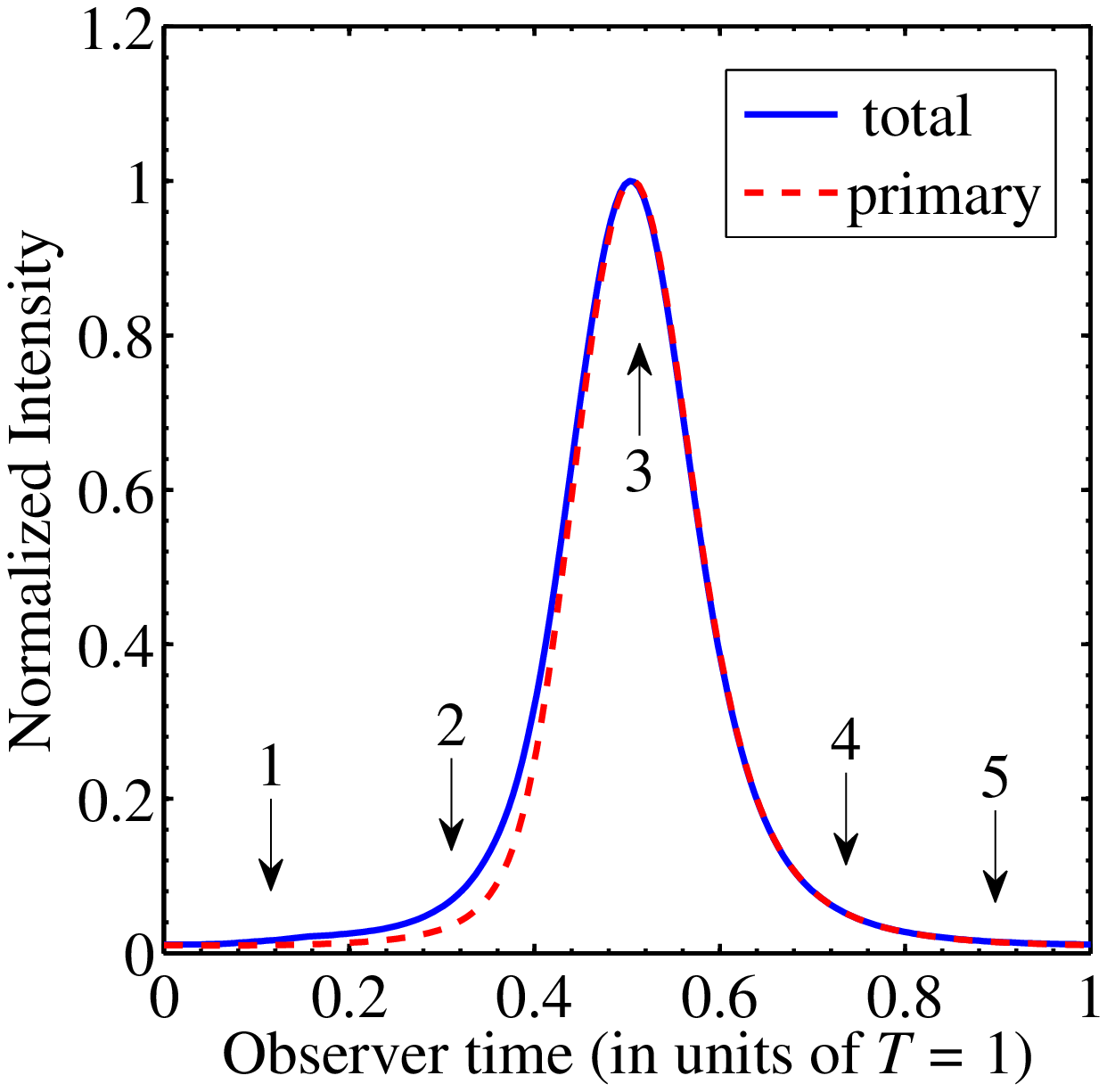}
\end{center}
\caption{Left panel: spectrogram of a hot spot with a radius of $R_{\rm spot} = 0.5 \, M$ 
orbiting a Schwarzschild BH at the ISCO. The observer's viewing angle is 
$i = 60^\circ$. Right panel: light curves of the same hot spot (blue solid line for
the total light curve, red dashed line for the light curve of the primary image only).
The five arrows refer to the five images shown in Figs.~\ref{fig1} and \ref{fig1bis}. 
See the text for more details.}
\label{fig2}
\end{figure}

\begin{figure}
\begin{center}
\includegraphics[height=7cm]{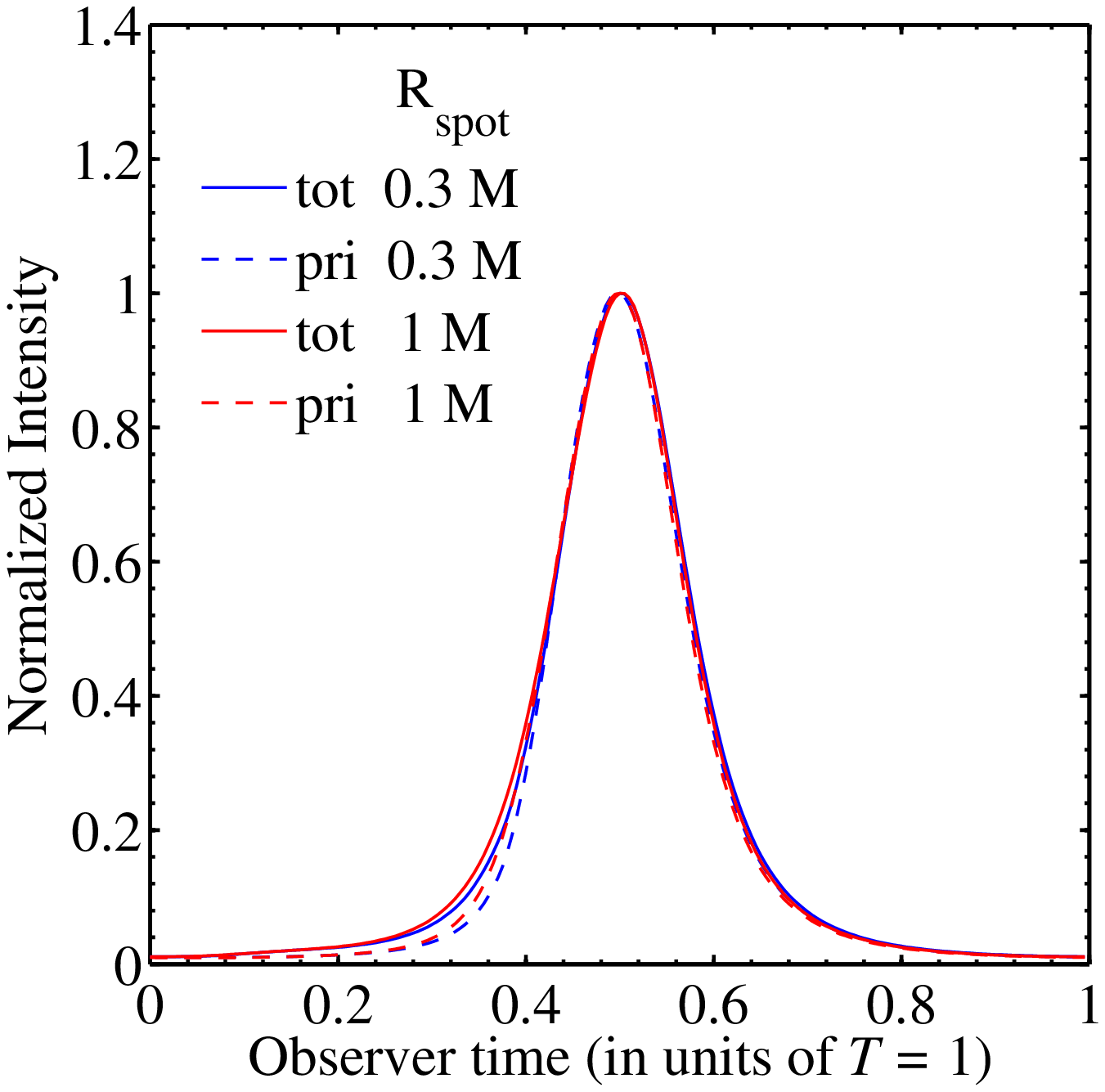}
\includegraphics[height=7cm]{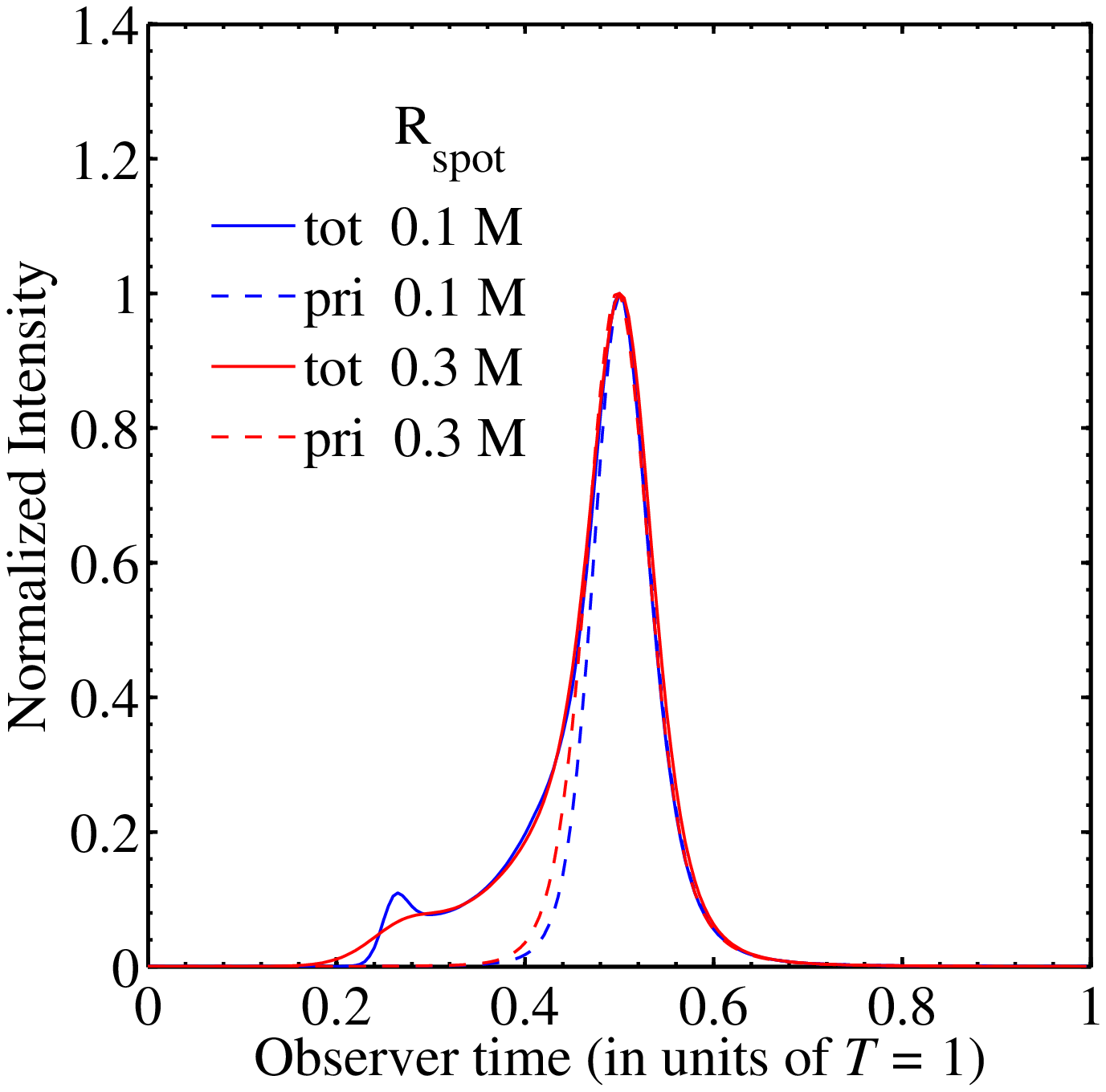}
\end{center}
\caption{Left panel: light curves of hot spots with a radius of $R_{\rm spot} = 0.3 \, M$ 
(blue/dark lines) and $1.0 \, M$ (red/light lines) orbiting a Schwarzschild BH at 
the ISCO. The observer's viewing angle is $i = 60^\circ$. Solid lines for the total 
light curves, dashed lines for the primary image light curves. Right panel: as in 
the left panel for hot spots with a radius of $R_{\rm spot} = 0.1 \, M$ (blue/dark lines) 
and $0.3 \, M$ (red/light lines) orbiting a Kerr BH with $a/M = 0.9$ at the ISCO. 
See the text for more details.}
\label{fig4}
\begin{center}
\includegraphics[height=7cm]{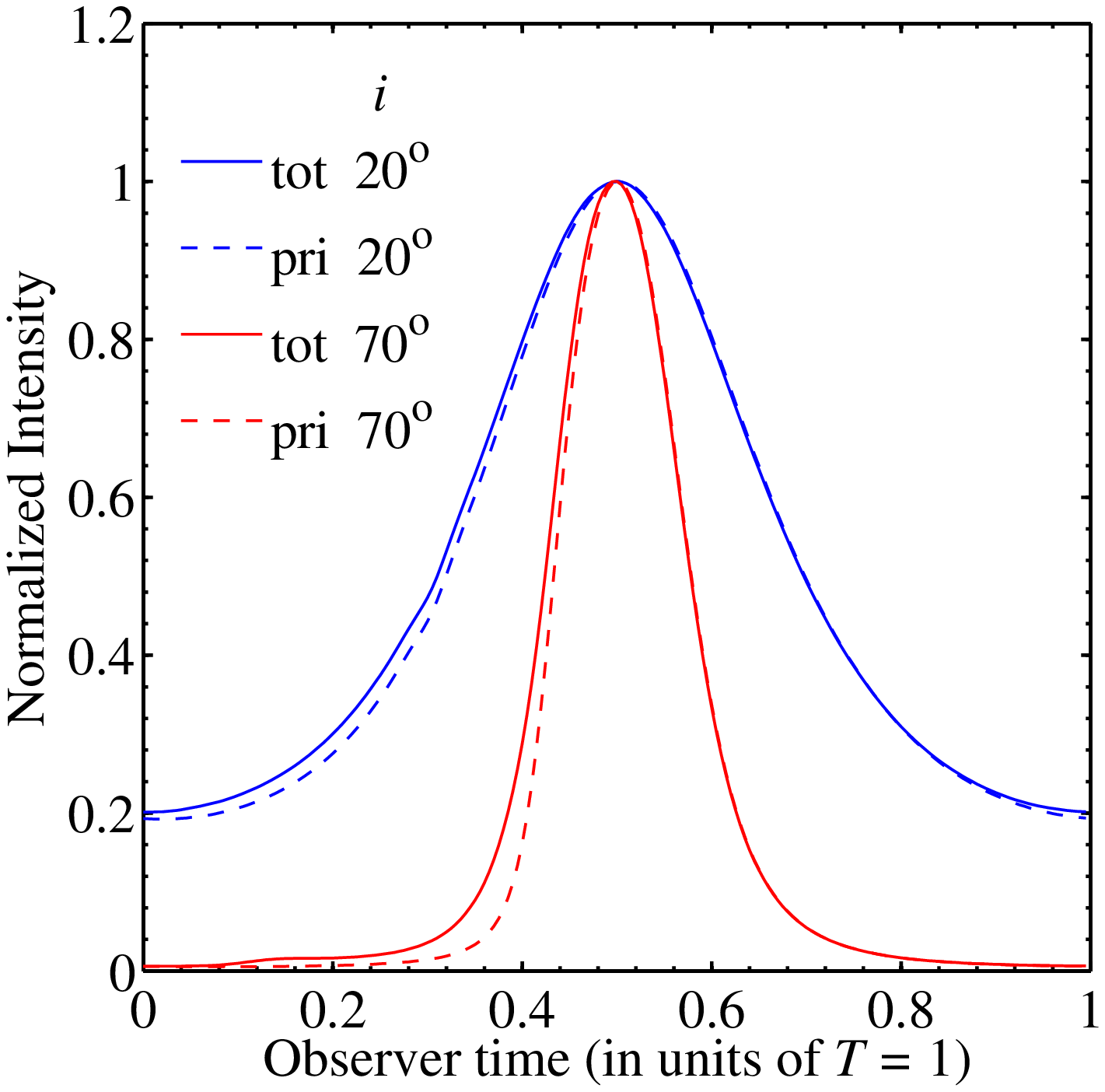}
\includegraphics[height=7cm]{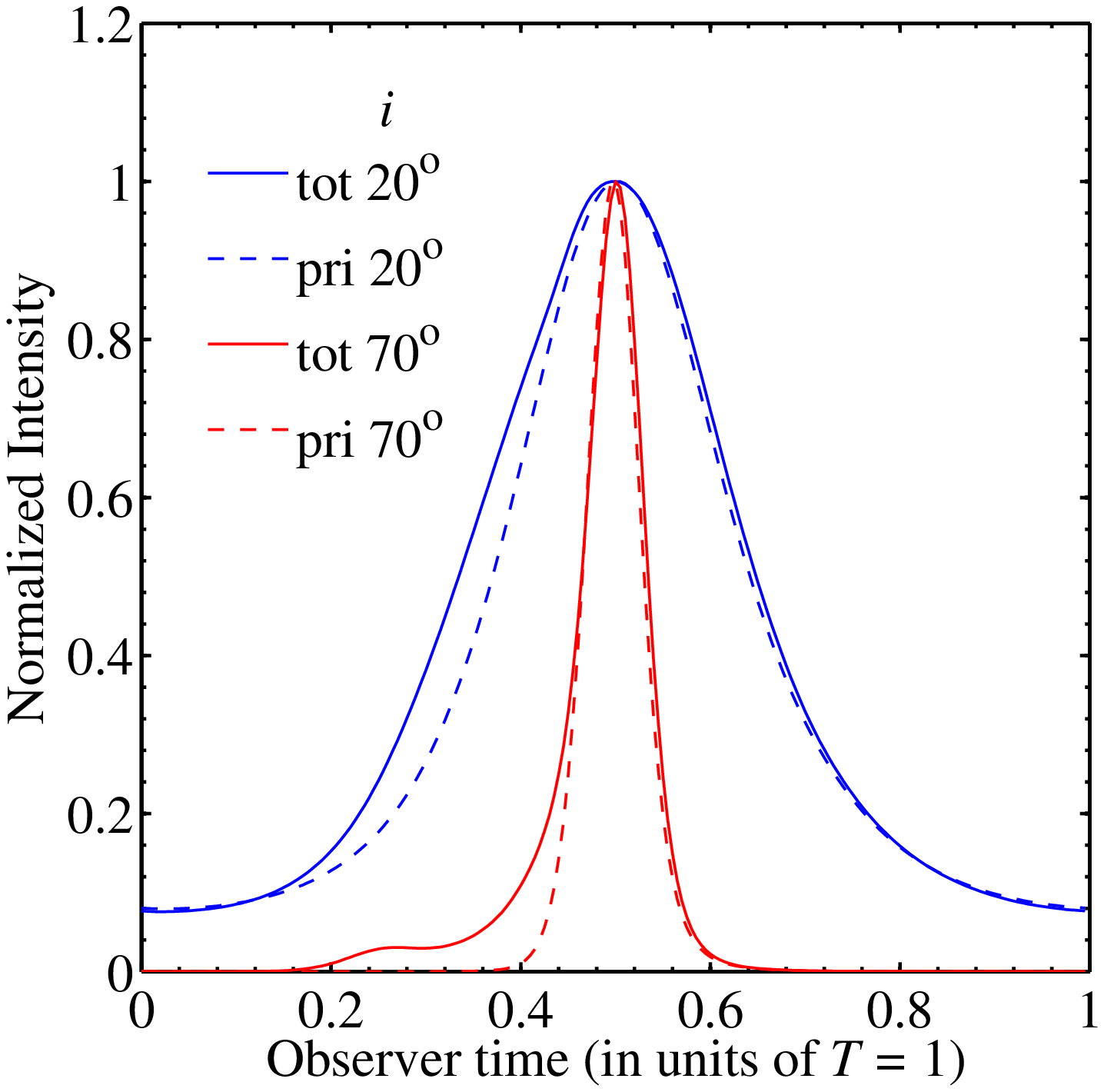}
\end{center}
\caption{Left panel: light curves of a hot spot with a radius of $R_{\rm spot} = 0.5 \, M$ 
orbiting a Schwarzschild BH at the ISCO. The observer's viewing angle is $i = 20^\circ$ 
(blue/dark lines) and $70^\circ$ (red/light lines). Solid lines for the total light curves, 
dashed lines for the primary image light curves. Right panel: as in the left panel for 
a hot spot with a radius of $R_{\rm spot} = 0.3 \, M$ orbiting a Kerr BH with $a/M = 0.9$ 
at the ISCO. See the text for more details.}
\label{fig5}
\vspace{1.0cm}
\end{figure}

\begin{figure}
\begin{center}
\includegraphics[height=7cm]{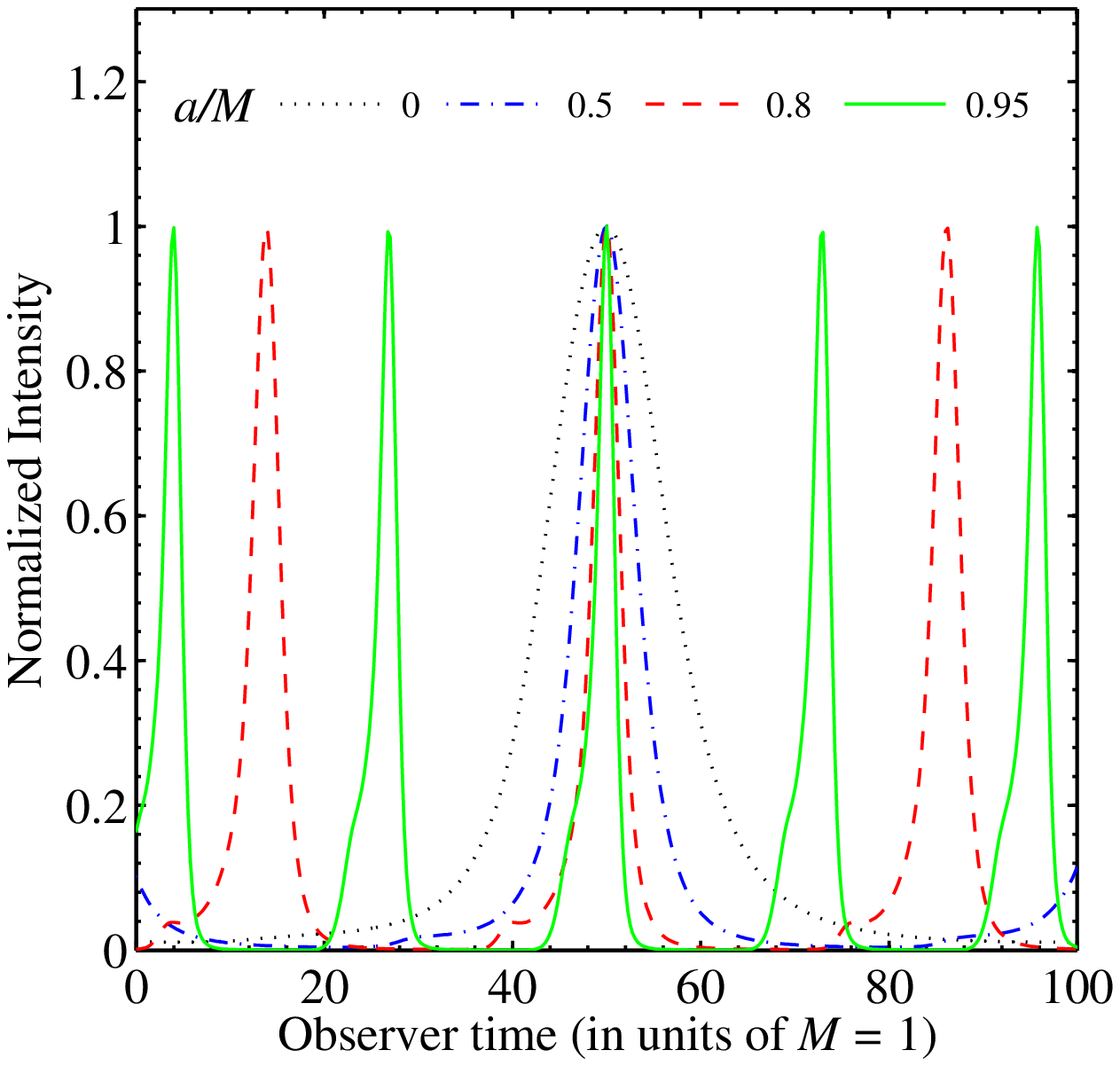}
\includegraphics[height=7cm]{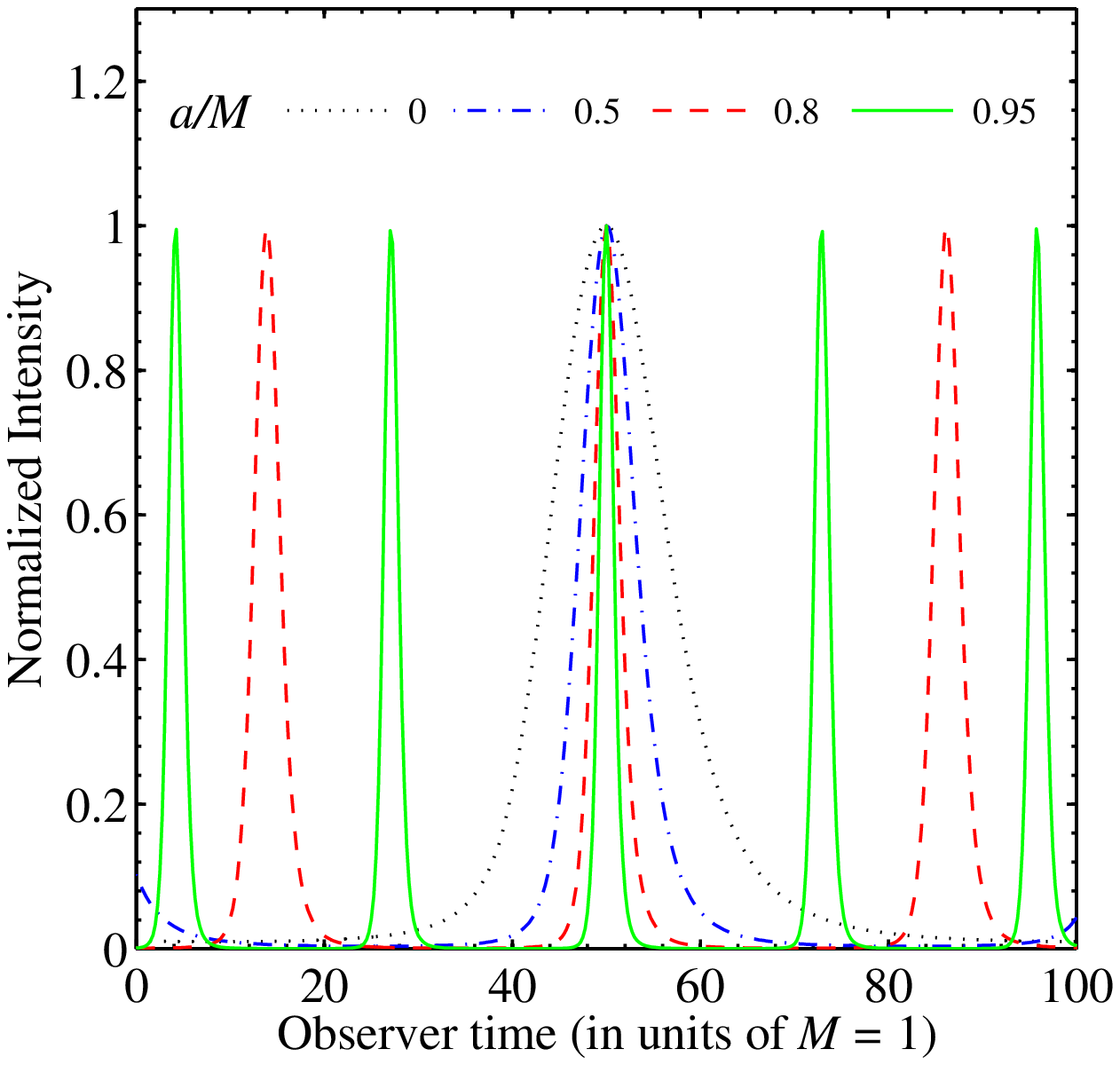} \\
\vspace{0.5cm}
\includegraphics[height=7cm]{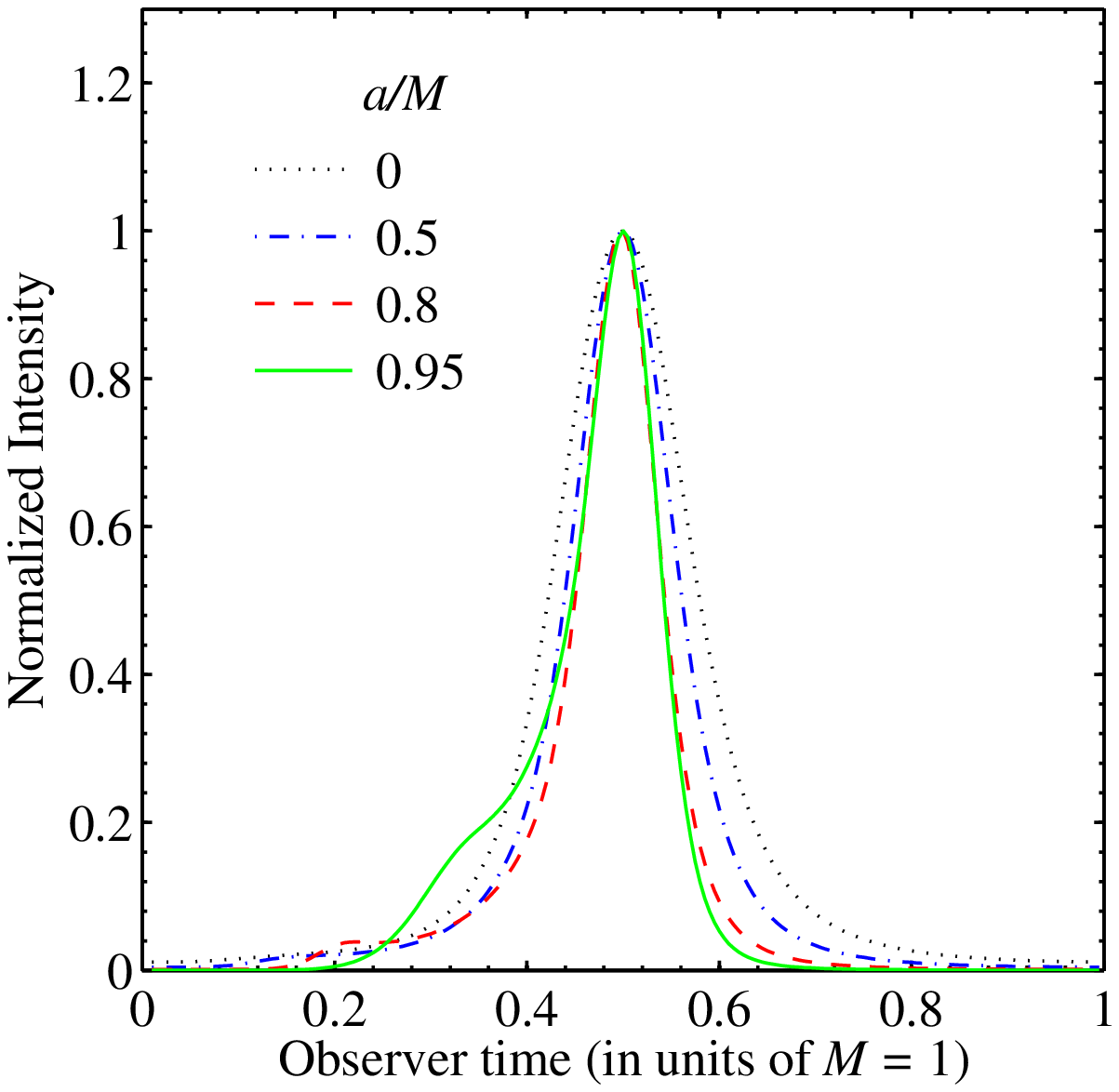}
\includegraphics[height=7cm]{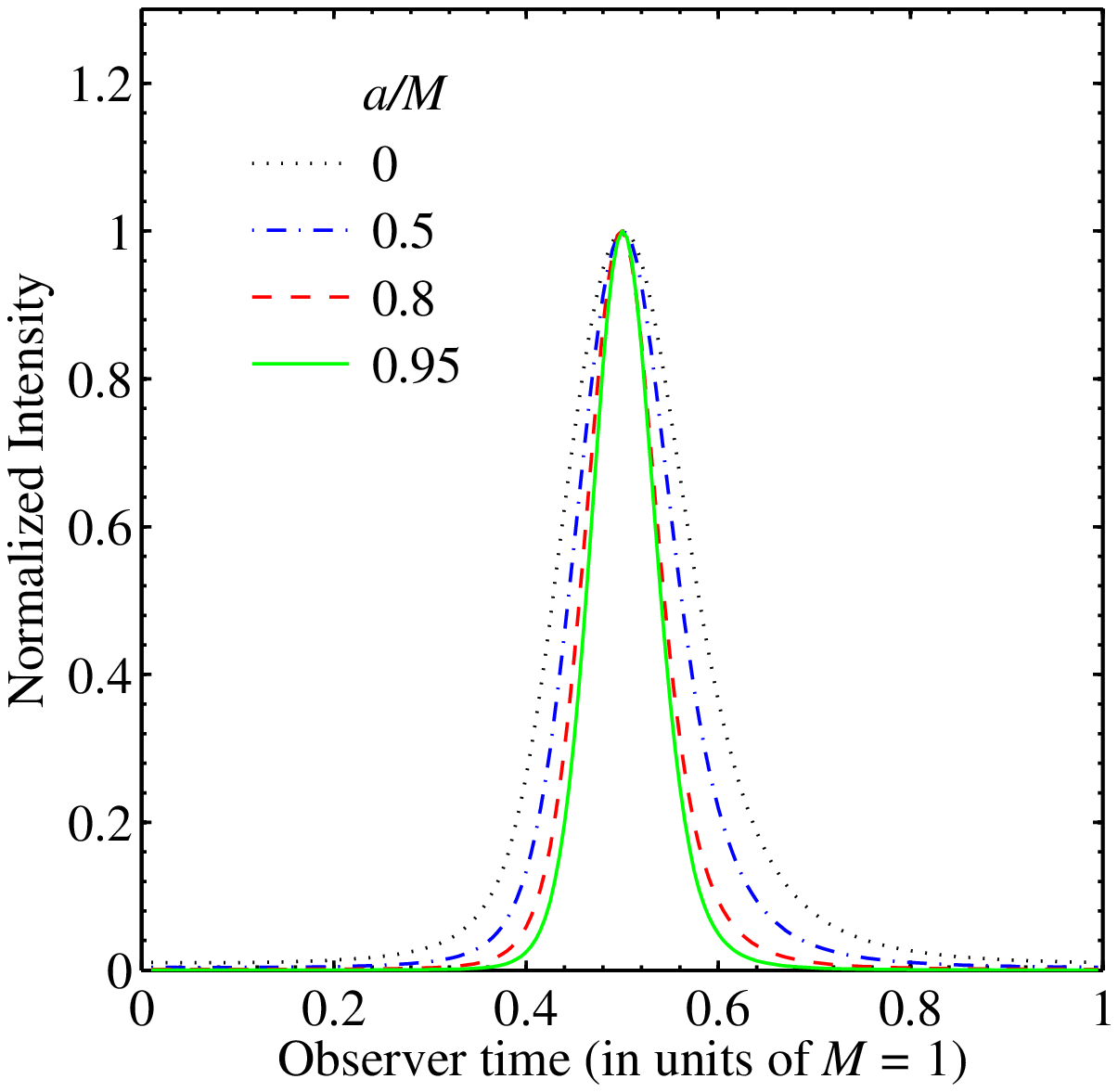}
\end{center}
\caption{Top panels: light curves (left) and primary image light curves (right) of
hot spots with a radius of $R_{\rm spot} = 0.3 \, M$ orbiting Kerr BHs with spin $a/M = 0$, 
0.5, 0.8 and 0.95 at the corresponding ISCO. The observer's viewing angle is 
$i = 60^\circ$. The frequency of the hot spot depends significantly on $a/M$, but 
an independent estimate of $M$ is necessary. Bottom panels: as in the top panels, 
but with the observer's time in units $T = 1$, where $T$ is the orbital period of 
the hot spot. See the text for more details.}
\label{fig5c}
\begin{center}
\includegraphics[height=7cm]{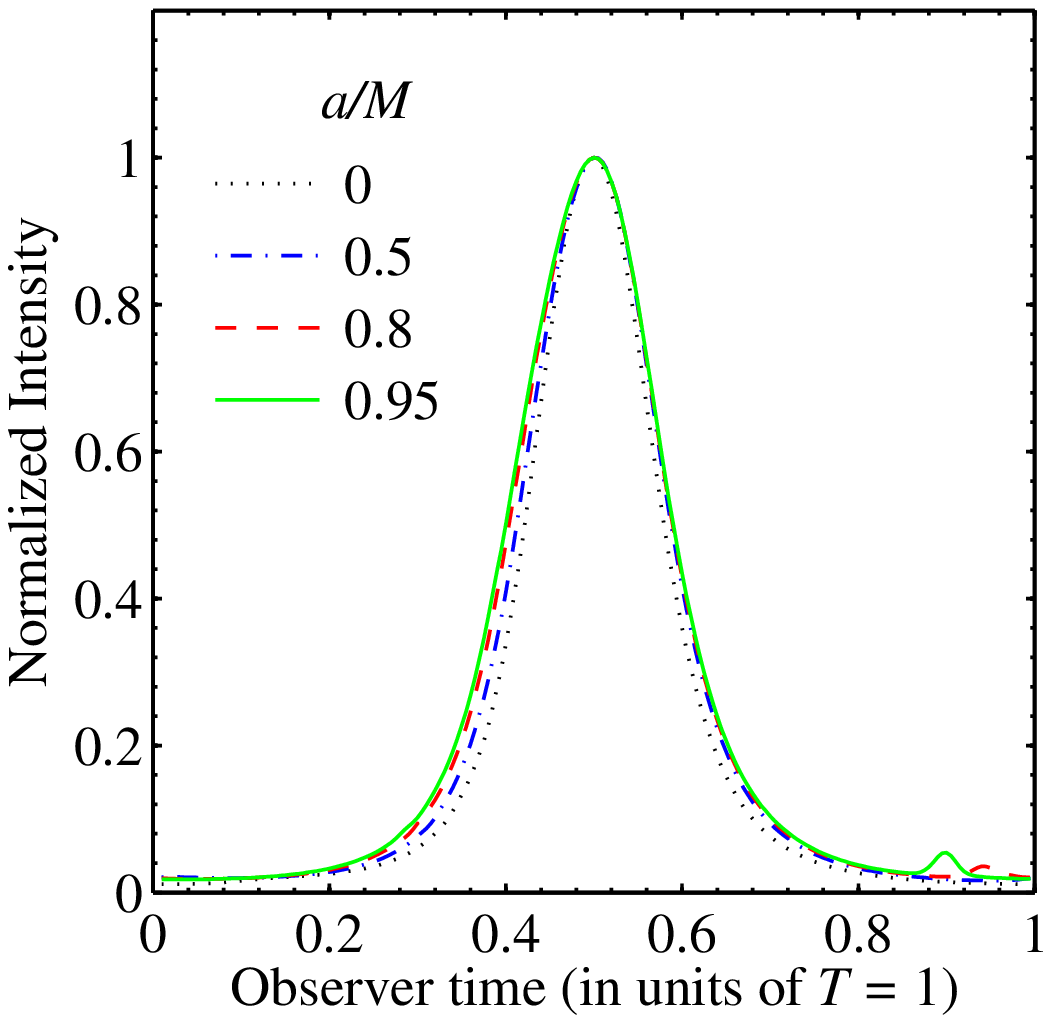}
\includegraphics[height=7cm]{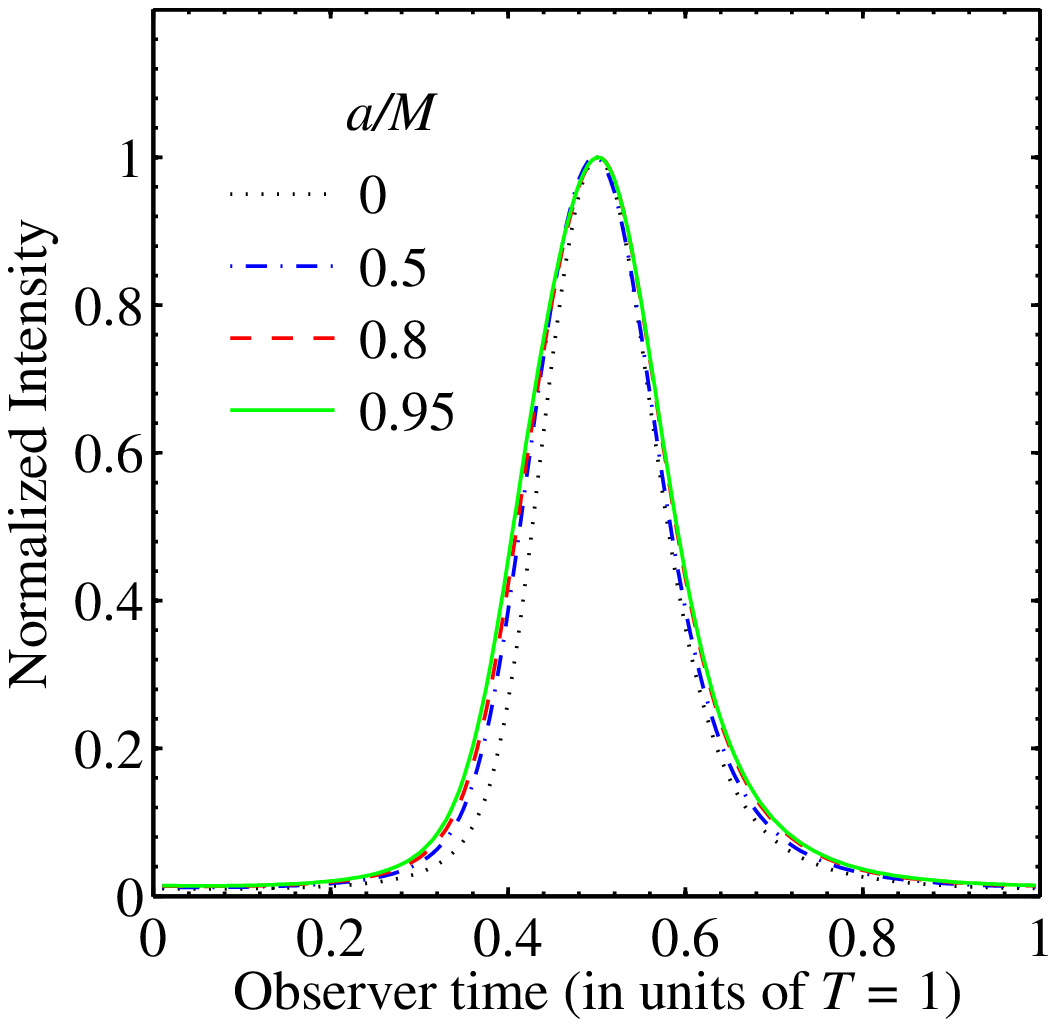}
\end{center}
\caption{Left panel: light curves of hot spots with a radius of $R_{\rm spot} = 0.3 \, M$ 
orbiting Kerr BHs with spin $a/M = 0$, 0.5, 0.8 and 0.95. For $a/M = 0$, the hot 
spot is at the ISCO, while in the other cases the hot spot is at the radius with same 
Keplerian angular frequency as the one orbiting a Schwarzschild BH at the ISCO. 
The observer's viewing angle is $i = 60^\circ$. Right panel: as in the left panel for 
the primary images only. See the text for more details.}
\label{fig5b}
\vspace{1.0cm}
\end{figure}

The photon trajectory is numerically integrated backward in time from the image 
plane of the distant observer to the point of the photon emission on the orbital
plane of the hot spot. We thus get the radial coordinate $r_{\rm e}$ at which the 
photon was emitted and the angle $\xi$ between the wavevector of the photon
and the normal of the orbital plane of the hot spot. The specific intensity of the 
radiation measured by the distant observer is given by
\be
I_{\rm obs}(\nu_{\rm obs},t_{\rm obs}) 
= g^3 I_{\rm em} (\nu_{\rm em},t_{\rm obs}) \, ,
\ee
where $g$ is the redshift factor
\be
g=\frac{E_{\rm obs}}{E_{\rm em}} = \frac{\nu_{\rm obs}}{\nu_{\rm em}} =
\frac{k_\alpha u^\alpha _{\rm obs}}{k_\beta u^\beta _{\rm em}} \, ,
\ee
$k_\alpha$ is the 4-momentum of the photon, $u_{\rm obs}^\alpha = (-1, 0, 0, 0)$ 
is the 4-velocity of the distant observer, and $u_{\rm em}^\alpha = (u_{\rm em}^t, 0, 0, 
\Omega u_{\rm em}^t)$ is the 4-velocity of the emitter. $\Omega$ is the Keplerian
angular frequency of a test-particle at the emission radius $r_{\rm e}$. 
$I_{\rm obs}(\nu_{\rm obs})/\nu^3_{\rm obs} = I_{\rm em}(\nu_{\rm em})/\nu^3_{\rm em}$ 
follows from the Liouville theorem. The hot spot emission is assumed to be 
monochromatic and isotropic, with a Gaussian intensity, as shown in Eq.~(\ref{emissivity}). 
Using the normalization condition $g_{\mu\nu}u^\mu_{\rm em}u^\nu_{\rm em} = -1$, 
one finds
\be
u_{\rm em}^t = -\frac{1}{\sqrt{-g_{tt}-2g_{t\phi}
\Omega-g_{\phi\phi}\Omega^2}} \, ,
\ee
and therefore,
\be
g = \frac{\sqrt{-g_{tt}-2g_{t\phi}\Omega-g_{\phi\phi}
\Omega^2}}{1+\lambda\Omega} \, , \label{redshift} 
\ee
where $\lambda = k_\phi/k_t$ is a constant of the motion along the photon path. 
Doppler boosting and gravitational redshift are entirely encoded in 
the redshift factor $g$. The effect of light bending is included by the raytracing 
calculation.

With the above machinery, we can compute the direct image of the hot spot at any
observer's time $t_{\rm obs}$. If the accretion disk around the BH is optically thick, 
as in the case of the one around stellar-mass BH candidates in binary systems in 
the X-ray band or of the accretion flow around SgrA$^*$ at mm wavelengths or
above, light rays do not pass through the disk (which is supposed to be on the 
equatorial plane) and therefore we do not see multiple images. On the other 
hand, if the accretion flow is optically thin, as it is expected for SgrA$^*$ at 
shorter wavelengths~\citep{shk}, light rays can pass through the disk and we can 
see multiple images of the hot spot, as a result of the gravitational lensing in the 
strong gravitational field around the object. Fig.~\ref{fig1} shows 5 instantaneous 
snapshots of a hot spot orbiting a Schwarzschild BH at the ISCO radius. These
images are observed at time intervals $T/5$, where $T$ is the orbital period of the
hot spot. The hot spot has a radius $R_{\rm spot} = 0.5 \, M$, it is moving 
anti-clockwise, and it is seen by an observer with viewing angle $i = 60^\circ$.
The color of the images represents the relative specific intensity of the radiation
(in arbitrary units). The left panel is for the primary images, the right panel for the
secondary ones. The secondary images are clearly dimmer and around the
apparent photon capture radius of the BH. The intensity of the center of the spot 
is always higher than the intensity of its edge because we are assuming the 
local specific intensity of Eq.~(\ref{emissivity}). Fig.~\ref{fig1bis} shows the redshift 
factor $g$ of the same images. With the GRAVITY instrument, one can expect 
to have NIR 10~$\mu$as astrometric measurements with a time resolution of 
about 1~minute~\citep{v2-model4}, to be compared with the hot spot orbital 
period $T$ of about 20~minutes. At radio and sub-mm wavelengths, integration 
times are significantly longer than in the NIR, and therefore radio and sub-mm 
facilities will not be able to detect these instantaneous images, but they will 
average them over the instrument time scale which is longer than the hot spot 
orbital period.

By integrating the observed specific intensity over the solid angle subtended by the 
image of the hot spot on the observer's sky, we obtain the observed flux
\be
F(\nu_{\rm obs},t_{\rm obs}) = 
\int I_{\rm obs}(\nu_{\rm obs},t_{\rm obs}) \, \mathrm{d}\Omega_{\rm obs} =
\int g^3 I_{\rm em}(\nu_{\rm em},t_{\rm obs}) \, \mathrm{d}\Omega_{\rm obs} \, . 
\ee
The complete spectrogram of the hot spot of Figs.~\ref{fig1} and \ref{fig1bis} is 
shown in the left panel of Fig.~\ref{fig2}, where the numbers 1, ..., 5 refer to the time 
of the images with the same number
in Figs.~\ref{fig1} and \ref{fig1bis}. For instance, the image numbered 
by 1 in Figs.~\ref{fig1} and \ref{fig1bis} corresponds to the observer's time $t/T=0.1$ in 
Fig.~\ref{fig2}. In this spectrogram, it is easy to identify the contribution coming from 
the primary image and the smaller one from the secondary image. 
If we integrate over the frequency range of the radiation, we get the observed 
luminosity, or light curve, of the hot spot
\be
L(t_{\rm obs}) = \int F(\nu_{\rm obs},t_{\rm obs}) \, \mathrm{d}\nu_{\rm obs} \, . 
\ee
The light curve of our hot spot is shown in the right panel of Fig.~\ref{fig2}, where
the blue/dark solid line is for the total light curve, while the red/light dashed line
shows the one from the primary image only. In the present paper, we normalize 
the light curves by dividing the observed luminosity $L(t_{\rm obs})$ by the
corresponding maximum, since only the shape of the light curve 
can be used to determine the parameters of the model.
Such a time-dependent emission signal can be added 
to a background intensity coming from the inner region of the steady state 
accretion disk. By definition, the hot spot will have a higher density and/or higher 
temperature and thus a higher emissivity than the background accretion disk, 
adding a small modulation to the total flux. For instance, in the case of stellar-mass 
BH candidates in X-ray binary systems, $RXTE$ observations find the high-frequency 
QPO modulations to have typical amplitudes of $1\%-5\%$ of the mean flux during 
the outburst~\citep{rem}.

At this point, we can see the dependence of the light curve on the model parameters.
The role of the size of the hot spot, its radius $R_{\rm spot}$, is outlined in Fig.~\ref{fig4}.
The left panel shows the light curve of a hot spot orbiting a Schwarzschild BH at the
ISCO; that is, at a radius $r = 6 \, M$ in Boyer-Lindquist coordinates. The size of the 
spot is clearly not very important, both in the total light curve and in the primary image 
one. The right panel of Fig.~\ref{fig4} shows the light curve of a hot spot orbiting at
the ISCO of a Kerr BH with $a/M = 0.9$. Now $r \approx 2.32 \, M$. The effect of the
spot size on the primary image light curve is larger than in the Schwarzschild case,
but still moderate. The secondary image light curve is instead much more sensitive 
to the value of $R_{\rm spot}$.

The effect of the inclination angle of the hot spot orbit with respect to the observer's
line of sight is shown in Fig.~\ref{fig5}. The left panel is again for a hot spot orbiting
at the ISCO radius around a Schwarzschild BH, while the right panel is for a Kerr
BH with $a/M = 0.9$. As the inclination increases, the light curve goes from nearly 
sinusoidal to being sharply peaked by the special relativistic effect of light beaming. 
As already pointed out in~\citet{brod1}, the value of the inclination angle determines
the magnification of the hot spot and therefore it can be potentially estimated from 
the measurement of the latter (but in real data the situation seems to be more 
complicated).

The effect of the spin is shown in Fig.~\ref{fig5c}. If the hot spot is orbiting at the 
ISCO and we know the mass $M$ of the BH, the spin determines the period of the
light curve, which changes significantly from $a/M = 0$ to $a/M = 1$. In the top
panels in Fig.~\ref{fig5c}, we chose a long enough time scale to include at least 
one period of each light curve, and we shifted one of the peaks of any light curve 
to the middle of the plot ($t=50 \, M$).
At least in principle, even if we did not know the mass $M$, the spin could be 
inferred from the sole shape of the light curve, as shown in the bottom panels 
of Fig.~\ref{fig5c}. With real data, this seems however to be unlikely.

If the hot spot is orbiting at a radius larger than the one of the ISCO, the orbital 
period increases, as well as the period of the light curve. A natural question is
if hot spots with the same orbital frequency but moving around Kerr BHs with 
different spin produce different light curves. The answer is partially positive: as 
shown in Fig.~\ref{fig5b} for the total light curve (left panel) and the primary 
image light curve (right panel), but the difference is quite small. It is interesting 
to note that, for moderate and high spins, the total light curve has a small second 
peak due to the secondary image. Even if the hot spot is at a relatively large 
radius, the secondary image is sensitive to the BH spin.
While the secondary maximum due to the secondary image is very low in the 
total light curve, and impossible to detect for present facilities like the ESO Very
Large Telescope (VLT), it can leave specific signatures in the centroid track,
which could be potentially observable for sufficiently bright spots and accurate
astrometric measurements [see Section~\ref{s-d} of the present 
paper or \citet{v2-model4}].

\vspace{1.0cm}

\section{Hot spots orbiting non-Kerr black holes}
\label{s-nk}

As shown in the previous section, images and light curves of hot spot orbiting
BHs are affected by a number of special and general relativistic effects, which
produce specific signatures in the observed electromagnetic radiation. If the
spacetime geometry around BH candidates is not described by the Kerr solution,
the predicted images and light curves of hot spots are presumably different,
and therefore their detection can potentially be used to test the Kerr nature
of an astrophysical BH candidate. In this section, we show how possible
deviations from the Kerr background can change the predictions of general
relativity, while in the next section we will present a more quantitative analysis,
paying some attention on the degeneracy/correlation between the estimate
of the BH spin and the constraint on possible deviations from the Kerr metric.

In order to test the Kerr metric around BH candidates it is useful to adopt
the following approach. We start considering a background metric more
general than the Kerr solution and that includes the Kerr metric as a special 
case. Here the compact object is characterized by a mass $M$, a spin 
parameter $a$, and one (or even more) deformation parameter(s) which 
measures possible deviations from the Kerr geometry. When the deformation 
parameter vanishes, we exactly recover the Kerr metric. We can then
compare the theoretical predictions obtained in this 3-parameter spacetime
with observational data. If the latter demand a vanishing deformation 
parameter, the Kerr BH hypothesis is verified; if we find a non-vanishing
deformation parameter, our data would suggest that the BH candidate is
not of the Kerr type. In general, however, it is not easy to arrive at a clear
conclusion, because most approaches are only sensitive to a particular
observable quantity that may be produced by a Kerr BH with a specific spin $a/M$,
as well as by many other non-Kerr BHs with different spin~\citep{cfm-iron}.
For instance, the continuum-fitting method actually measures something 
like the radiative efficiency in the Novikov-Thorne model, $\eta_{\rm NT} = 
1 - E_{\rm ISCO}$, where $E_{\rm ISCO}$ is the specific energy of a test 
particle at the ISCO. In the Kerr background, there is a one-to-one
correspondence between $\eta_{\rm NT}$ and $a/M$, and therefore this
technique can be used to estimate the spin parameter of the BH. However,
if we also have a deformation parameter, $\eta_{\rm NT}$ depends on both
the spin and the deformation parameter. If we fix the deformation parameter,
the same value of $\eta_{\rm NT}$ is found for a particular value of the 
spin, at least if the deformation parameter is not too large.
Such a degeneracy/strong correlation between the spin and the deformation
parameter can be fixed either by combining two measurements sensitive
to very different properties of the background metric~\citep{jet1,jet2,qpo2}, 
or by an observable quantity that is not sensitive to only one number.

\begin{figure}
\begin{center}
\includegraphics[height=7cm]{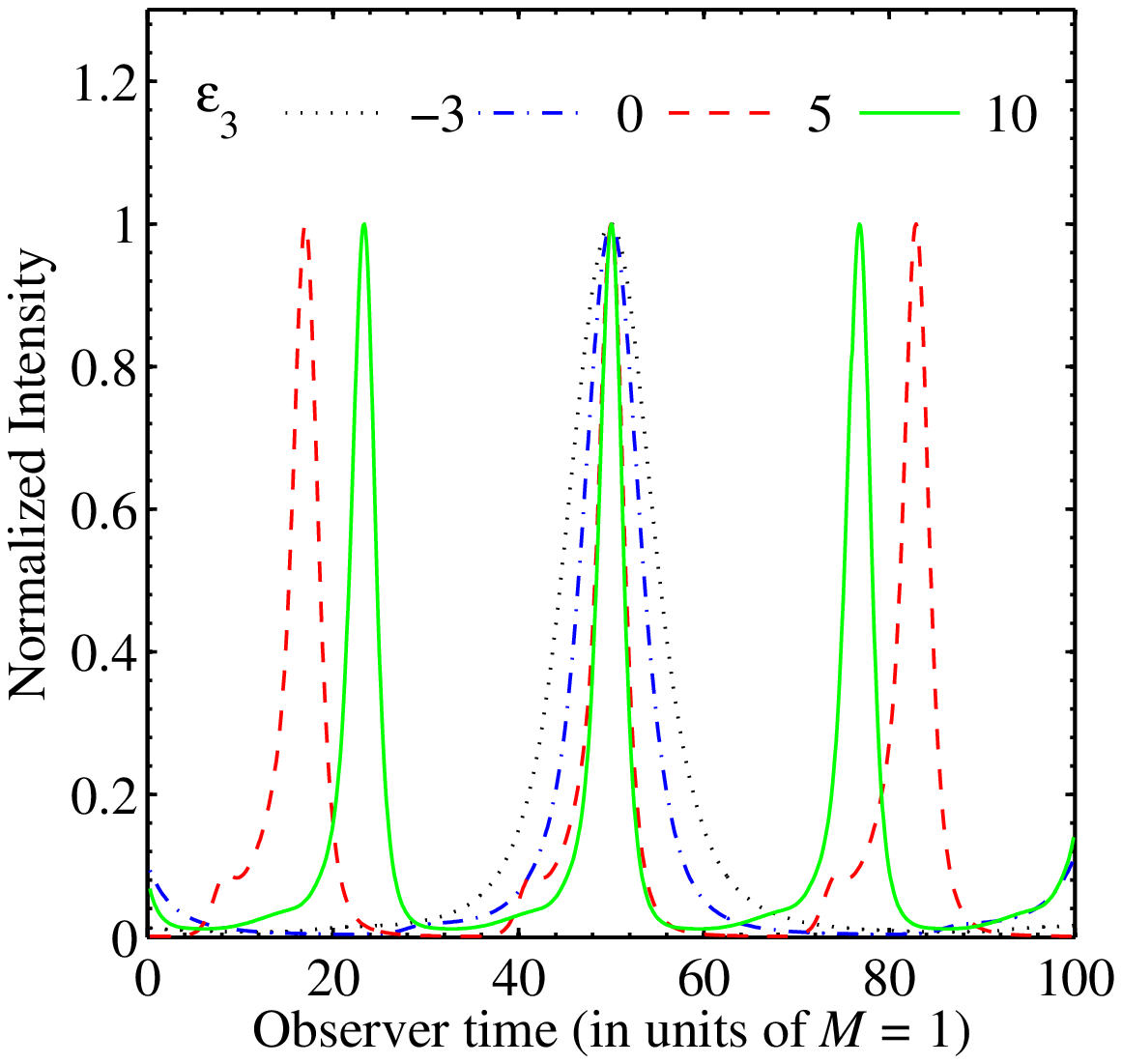}
\includegraphics[height=7cm]{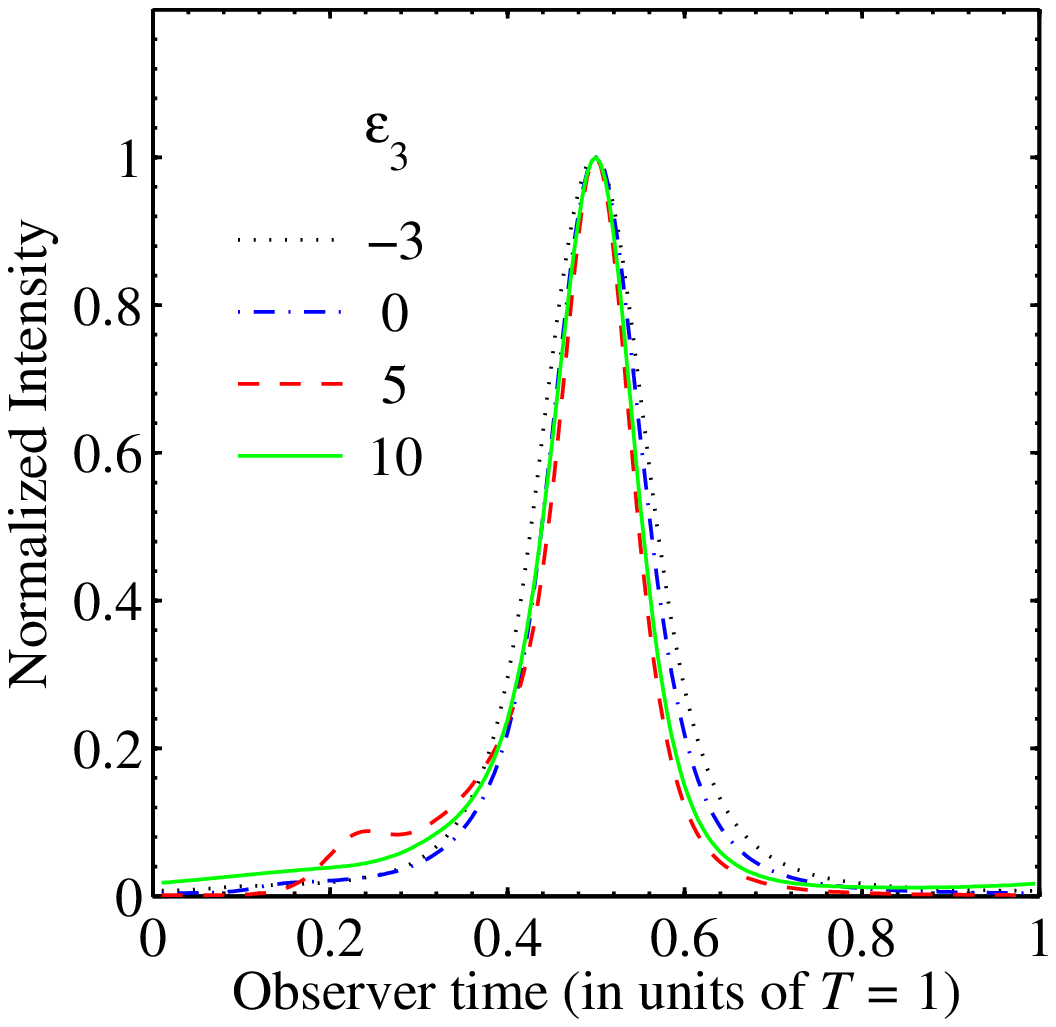}
\end{center}
\caption{Left panel: light curves of hot spots with a radius of $R_{\rm spot} = 0.3 \, M$
orbiting Johannsen-Psaltis BHs with $a/M = 0.5$ and $\epsilon_3 = -3$, 0, 5, 10 
at the ISCO. The observer's viewing angle is $i = 60^\circ$. Right panel: as in the 
left panel, but with the observer's time in units $T = 1$, where $T$ is the orbital 
period of the hot spot. See the text for more details.}
\label{fig6}
\begin{center}
\includegraphics[height=7cm]{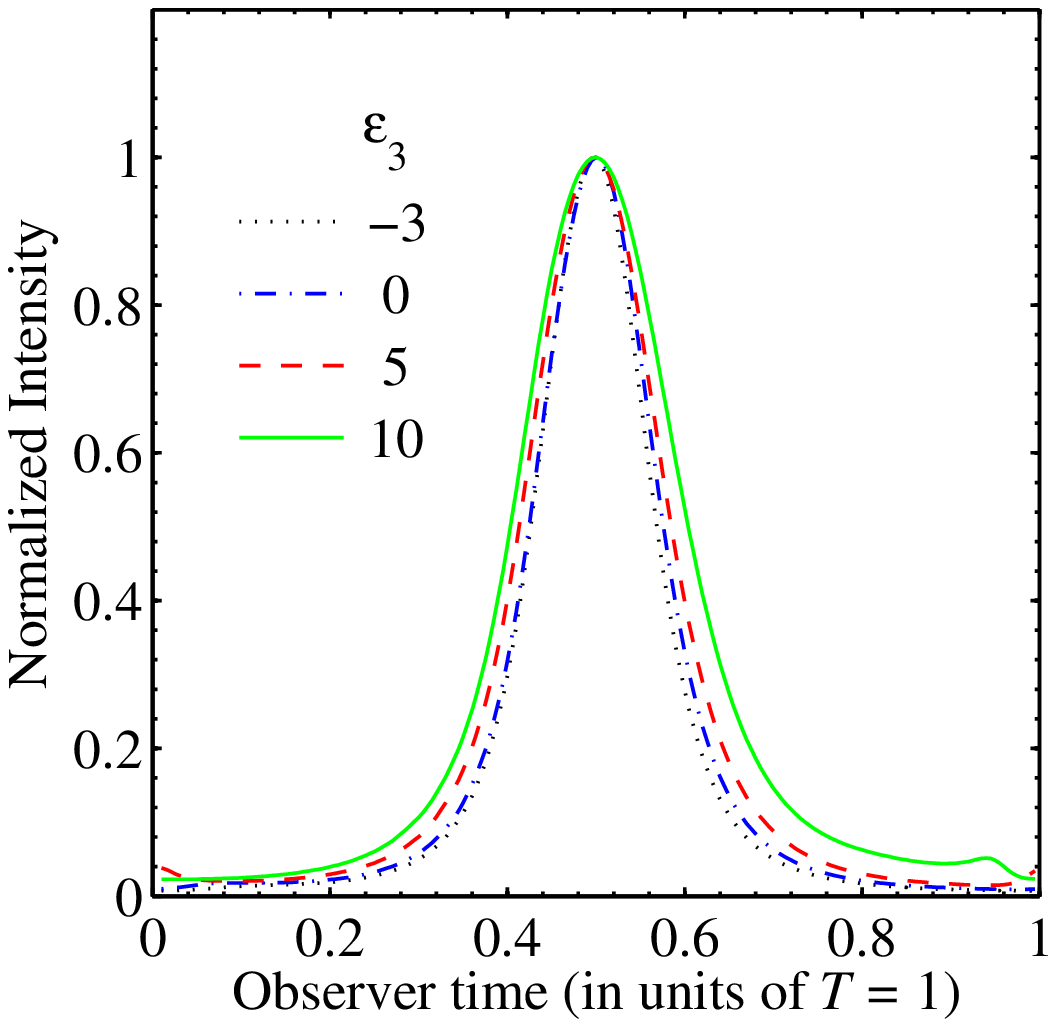}
\includegraphics[height=7cm]{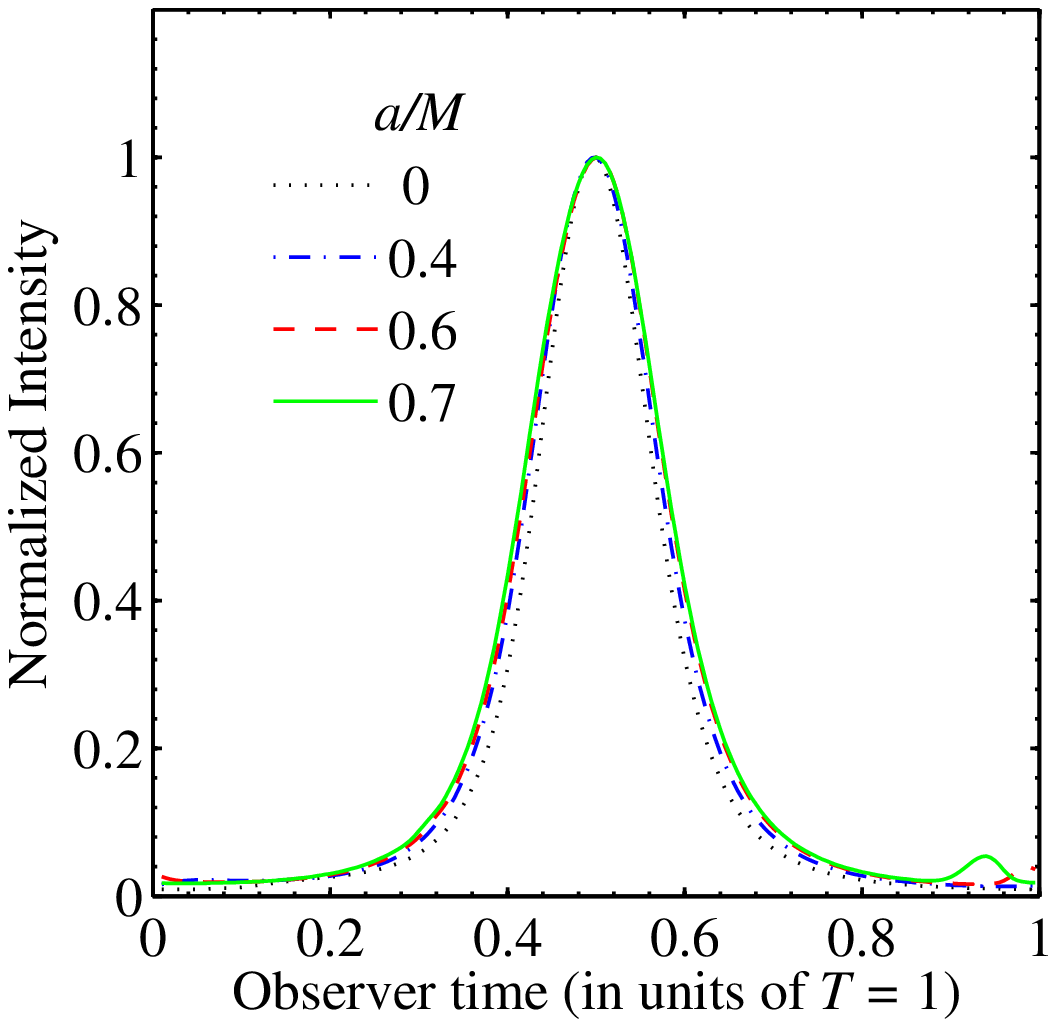}
\end{center}
\caption{Left panel: light curves of hot spots with a radius of $R_{\rm spot} = 0.3 \, M$ 
orbiting Johannsen-Psaltis BHs with $a/M = 0.5$ and $\epsilon_3 = -3$, 0, 5, 10.
The hot spot is at the radius with the same angular frequency as the one at the
ISCO of the BH with $\epsilon_3 = -3$. The observer's viewing angle is $i = 60^\circ$. 
Right panel: light curves of hot spots with a radius of $R_{\rm spot} = 0.3 \, M$ orbiting
Johannsen-Psaltis BHs with $a/M = 0.0$, 0.4, 0.6, 0.7 and $\epsilon_3 = 3$. 
The hot spot is at the radius with the same angular frequency as the one at the 
ISCO of the BH with $a/M = 0$. See the text for more details.}
\label{fig7}
\vspace{1.0cm}
\end{figure}

As non-Kerr background, here we consider the Johannsen-Psaltis metric~\citep{metric}.
While the latter is not a solution of any known gravity theory, and for some values
of the background parameters it presents several pathological features, it is a simple
metric that parametrizes possible deviations from the Kerr geometry. In most cases,
observations can only check if the gravitational force around a BH with spin parameter 
$a/M$ is stronger or weaker than the one around a Kerr BH with the same spin, 
without being able to probe the unphysical region with pathological features. In 
Boyer-Lindquist coordinates, the metric is given by the line element~\citep{metric}
\begin{align}
\mathrm{d}s^2=&-\bigg(1-\frac{2Mr}{\Sigma}\bigg)(1+h)\mathrm{d}t^2+
\frac{\Sigma(1+h)}{\Delta+a^2h\sin^2\theta}\mathrm{d}r^2   
+\Sigma\mathrm{d}\theta^2-\frac{4aMr\sin^2\theta}{\Sigma}(1+h)\mathrm{d}t\mathrm{d}\phi  \nonumber \\
&+\bigg[\sin^2\theta\bigg(r^2+a^2+\frac{2a^2Mr\sin^2\theta}{\Sigma}\bigg)   
+\frac{a^2(\Sigma+2Mr)\sin^4\theta}{\Sigma}h\bigg]\mathrm{d}\phi^2,
\end{align}
where $\Sigma=r^2+a^2\cos^2\theta$, $\Delta=r^2-2Mr+a^2$, and
\be
h=\sum_{k=0}^\infty\bigg(\epsilon_{2k}+\frac{Mr}{\Sigma}
\epsilon_{2k+1}\bigg)\bigg(\frac{M^2}{\Sigma}\bigg)^k.
\ee
This metric has an infinite number of deformation parameters $\epsilon_i$, 
and the Kerr solution is recovered when all the deformation parameters are 
set to zero. However, in order to reproduce the correct Newtonian limit, we 
have to impose $\epsilon_0=\epsilon_1=0$, while $\epsilon_2$ is strongly 
constrained by Solar System experiments~\citep{metric}. In this paper, we 
will only examine the simplest cases where $\epsilon_3\neq0$, while all the 
other deformation parameters vanish. The choice of another deformation
parameter would not qualitatively change our results and conclusions. 
Negative $\epsilon_i$s make always the BH more oblate and therefore the 
gravitational force on the equatorial plane is stronger, with the result that the 
ISCO radius is larger and the ISCO orbital period longer. Positive $\epsilon_i$s 
have the opposite effect, but the ISCO radius and the ISCO period first
decrease and then increase as $\epsilon_i$ increases. The qualitative
effect of any deformation parameter is the same, see e.g.~\citet{v2-io}.
The presence of more than one non-vanishing deformation parameter
makes the discussion more complicated, but it does not introduce any new 
feature and for this reason we can restrict the attention to $\epsilon_3$.

Figs.~\ref{fig6} and \ref{fig7} show the effect of the deformation parameter 
$\epsilon_3$ on the (total) light curve of a hot spot. In Fig.~\ref{fig6}, we assume 
that the hot spot is at the ISCO. Here, the major effect is the variation of the
ISCO radius, which implies a different hot spot orbital period (left panel). 
However, an independent estimate of the mass $M$, which sets the characteristic
scale of the system, must be known. If we consider the case in which we do not know
at all $M$ (this is not the case of SgrA$^*$, even if at present the uncertainty 
on its measurement is quite large, around 10\%), we can compare light curves
over an orbital period, as shown in the right panel in Fig.~\ref{fig6}. A different 
deformation parameter still produces a light curve with a slightly different shape. 
In the end, if the hot spot is at the ISCO, the deformation parameter plays 
more or less the same role as the spin $a/M$, discussed in Fig.~\ref{fig5c}.

\begin{figure*}
\begin{center}
\includegraphics[height=7cm]{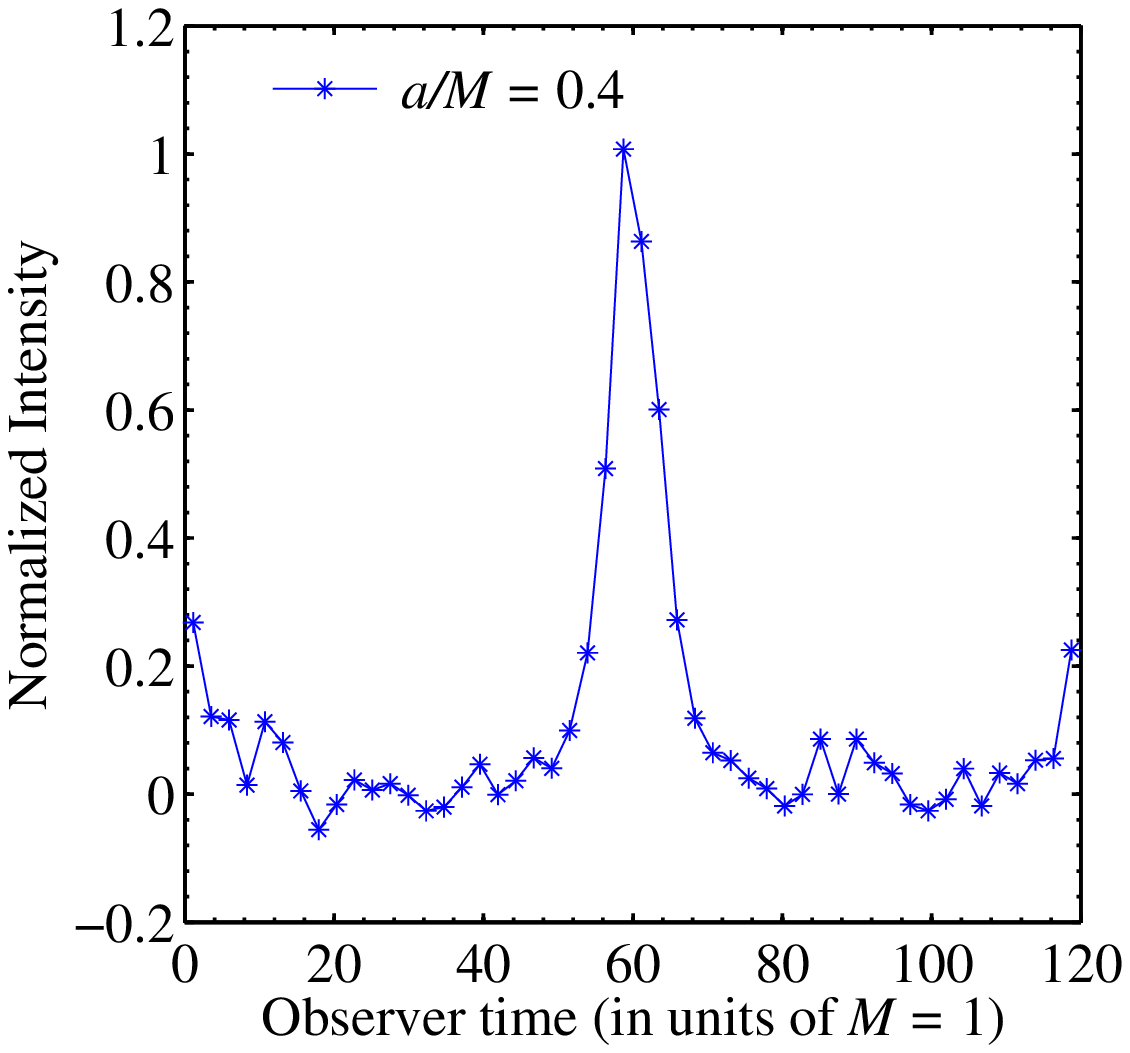}
\end{center}
\caption{Simulated total light curve of a hot spot orbiting the ISCO of a Kerr BH 
with $\tilde{a}/M = 0.4$. \label{fig13}}
\begin{center}
\includegraphics[height=7cm]{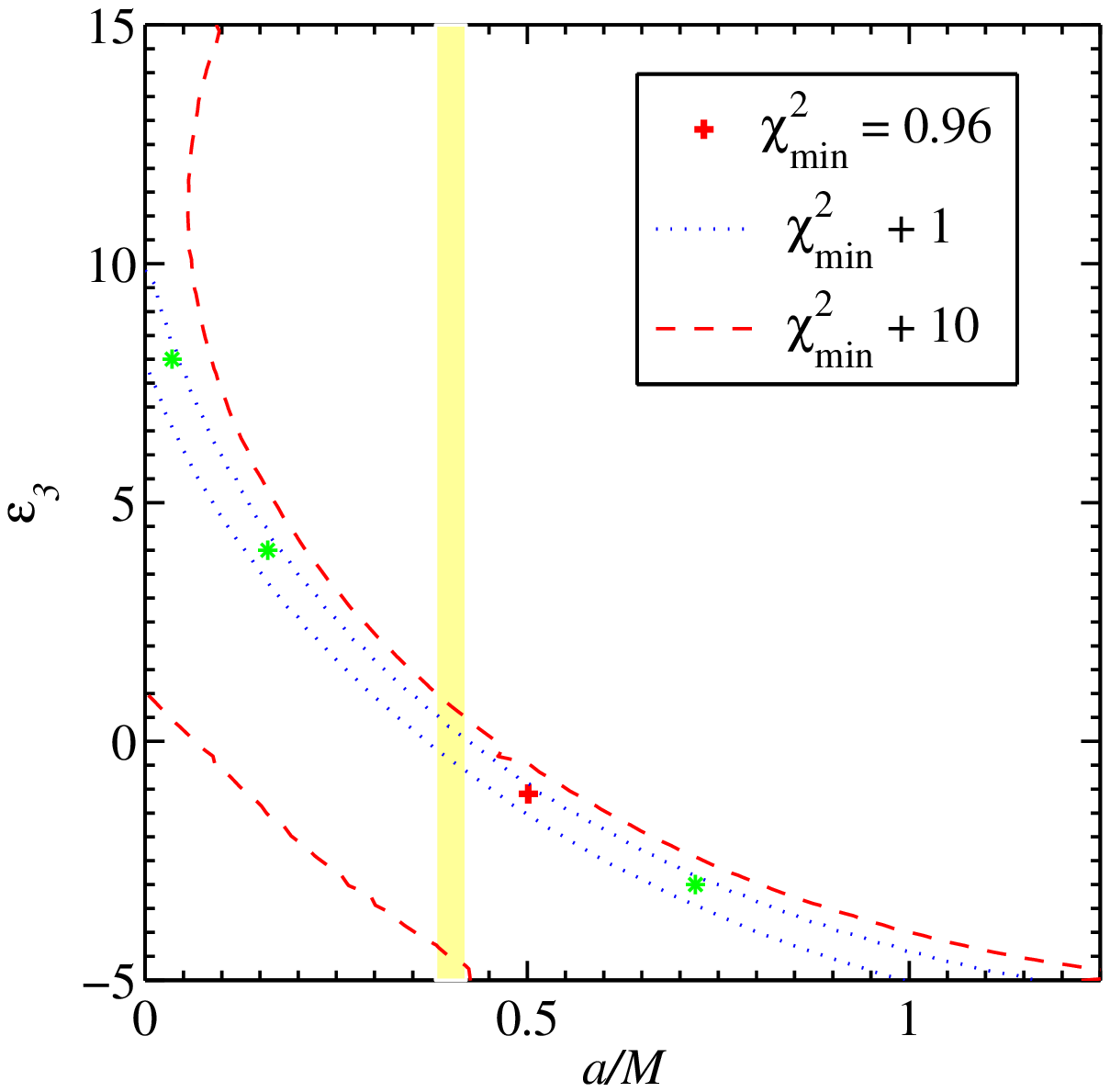}
\includegraphics[height=7cm]{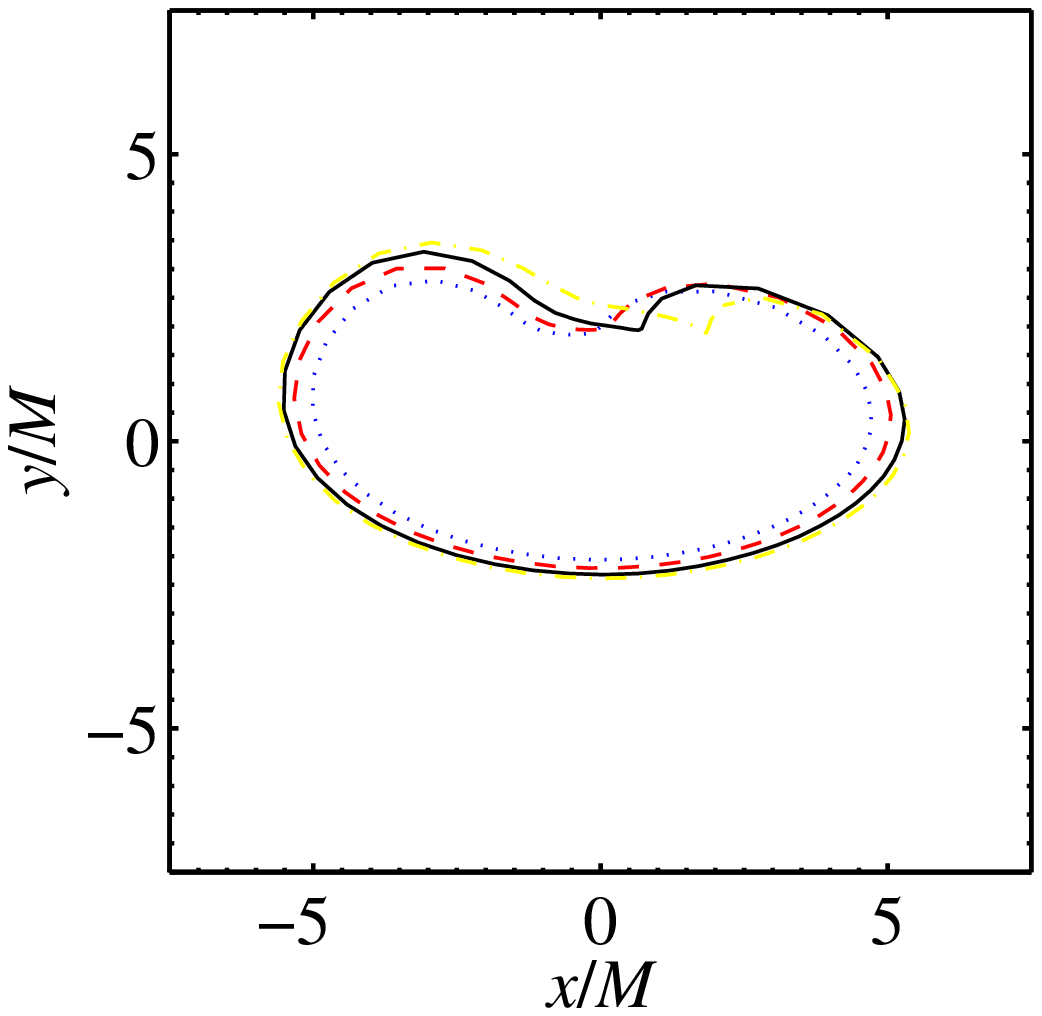}\\
\includegraphics[height=7cm]{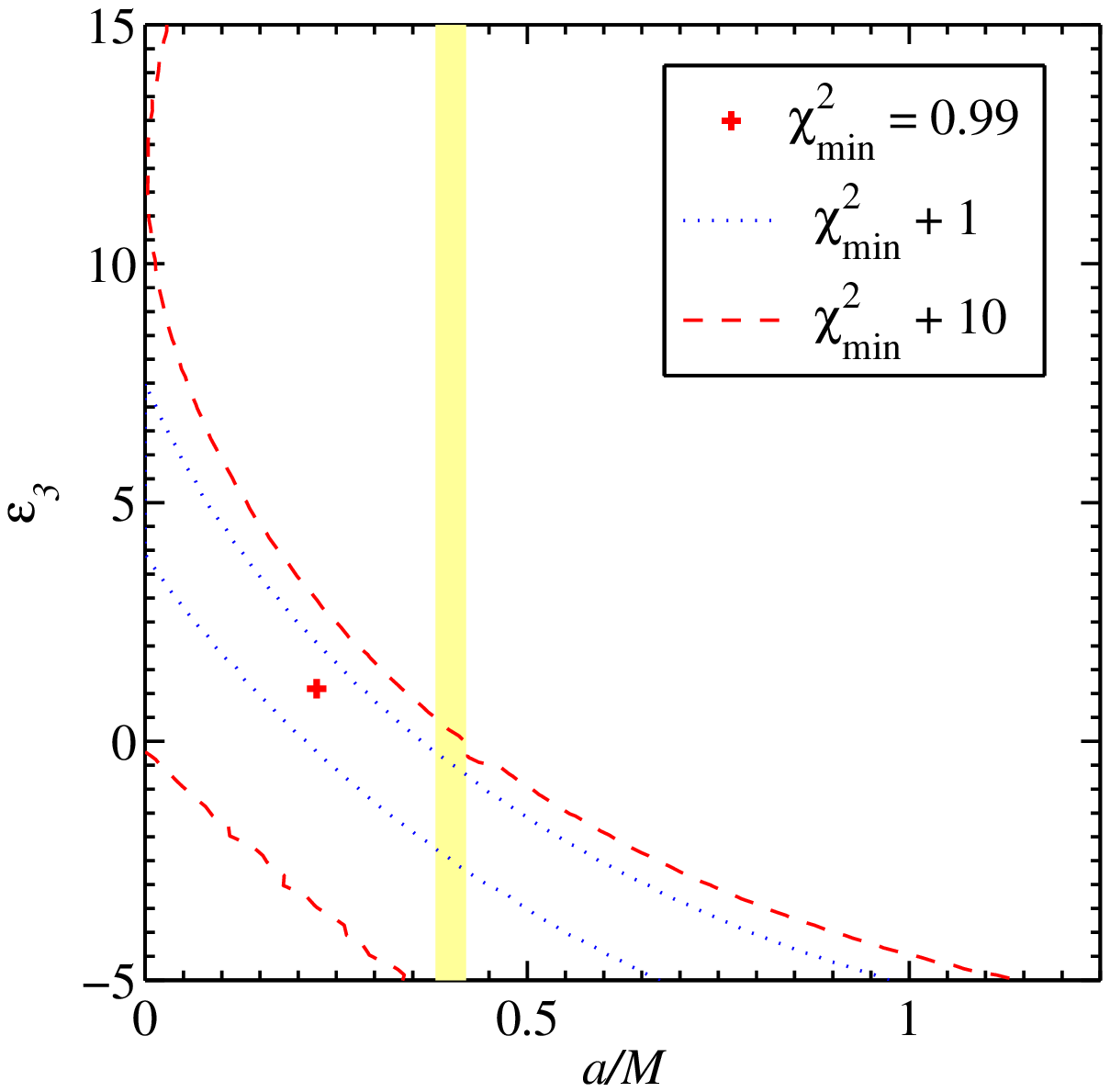}
\includegraphics[height=7cm]{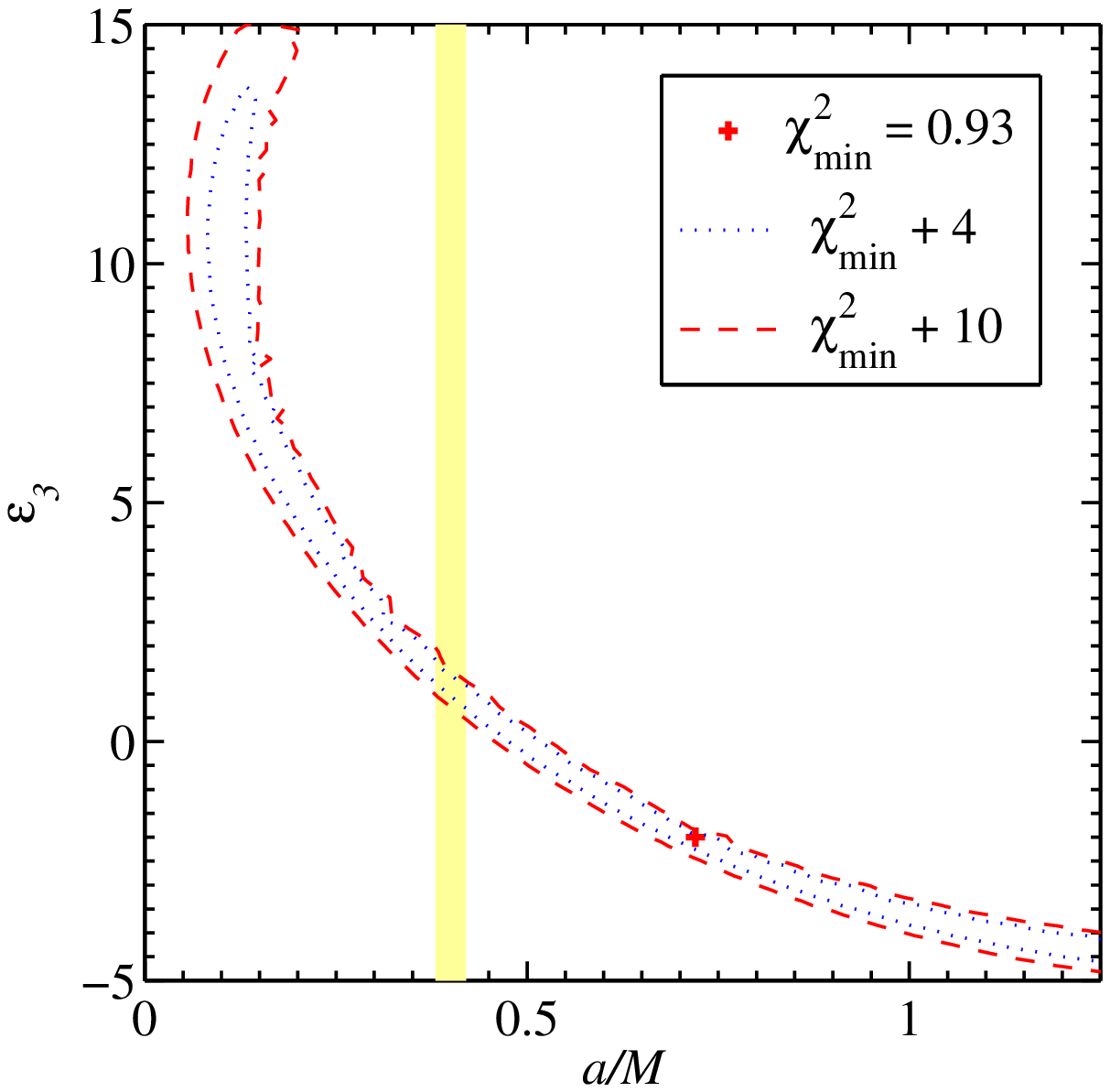}
\end{center}
\caption{Top left panel: $\chi^2_{\rm red}$ from the comparison of the simulated total 
light curve of a hot spot orbiting the ISCO of a Kerr BH with $\tilde{a}/M = 0.4$ (shown in 
Fig.~\ref{fig13}) and of the theoretical light curves expected from hot spots orbiting 
the ISCO of Johannsen-Psaltis BHs. Light curves from hot spots in spacetimes with
the same ISCO frequency are eventually indistinguishable. The hot spot size is 
$R_{\rm spot} = 0.3 \, M$ and the observer's viewing angle is $i = 60^\circ$. The 
yellow area corresponds to a possible BH spin measurement $a/M = 0.40 \pm 0.01$ 
from a radio pulsar in a compact orbit in the case in which SgrA$^*$ has $a/M = 0.4$ 
(such a measurement would be independent of $\epsilon_3$). Top right panel: 
centroid tracks of hot spots orbiting the ISCO of BHs with $a/M=0.04$ and 
$\epsilon_3=8$ (blue dotted curve), $a/M=0.16$ and $\epsilon_3=4$ (red dashed 
curve), $a/M=0.4$ and $\epsilon_3=0$ (black solid curve), and $a/M=0.72$ and 
$\epsilon_3=-3$ (yellow dashed-dotted curve). These four spacetimes correspond 
to the one of the simulated light curve and to the ones associated to the three green 
stars in the left panel. Bottom panels: as in the top left panel in the case of simulated 
light curves in a Kerr spacetime with $\tilde{a}/M = 0.3$ (right panel) and 0.5 (left panel).
See the text for more details. \label{fig14}}
\end{figure*}

\begin{figure*}
\begin{center}
\includegraphics[height=7cm]{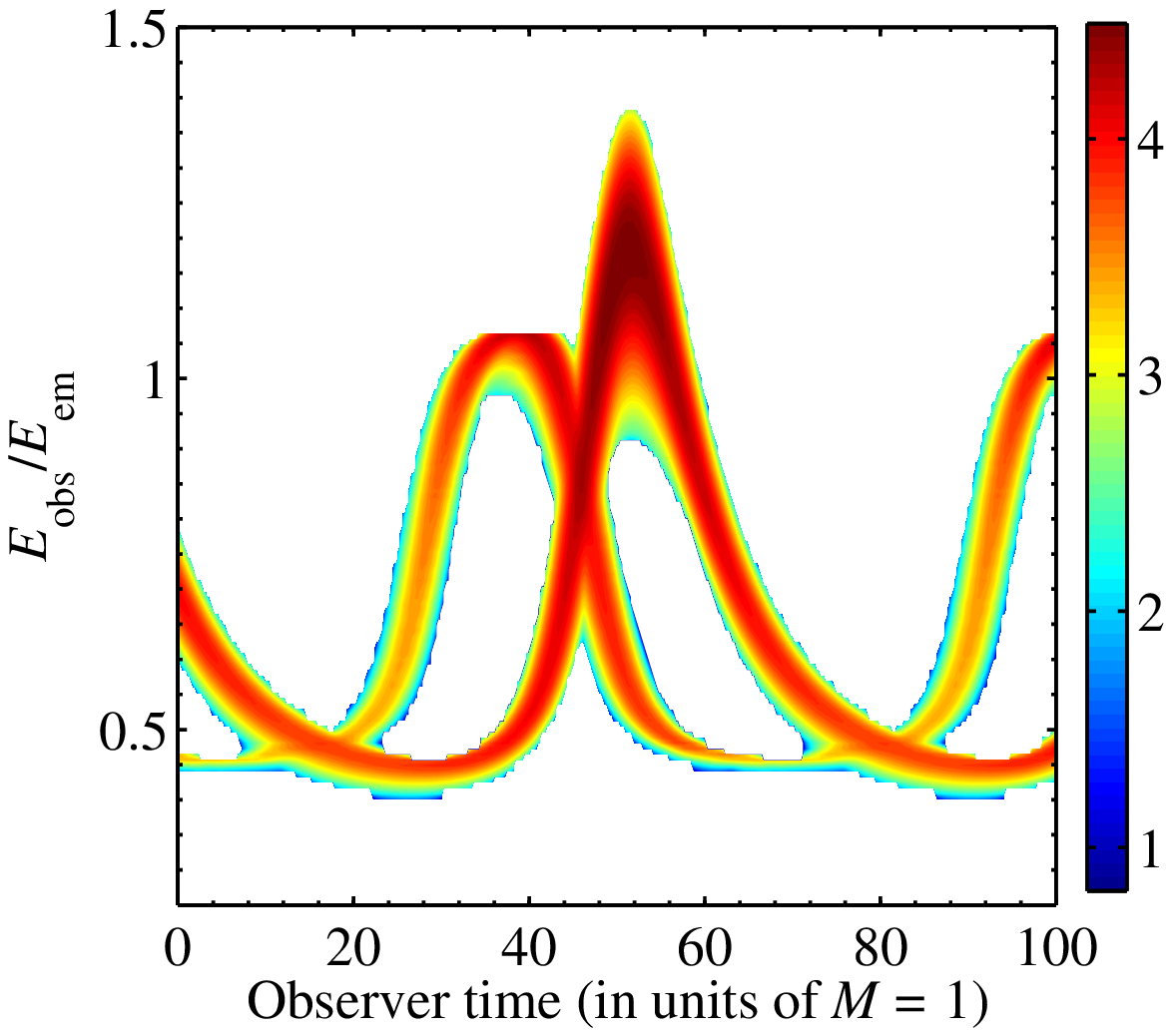}
\includegraphics[height=7cm]{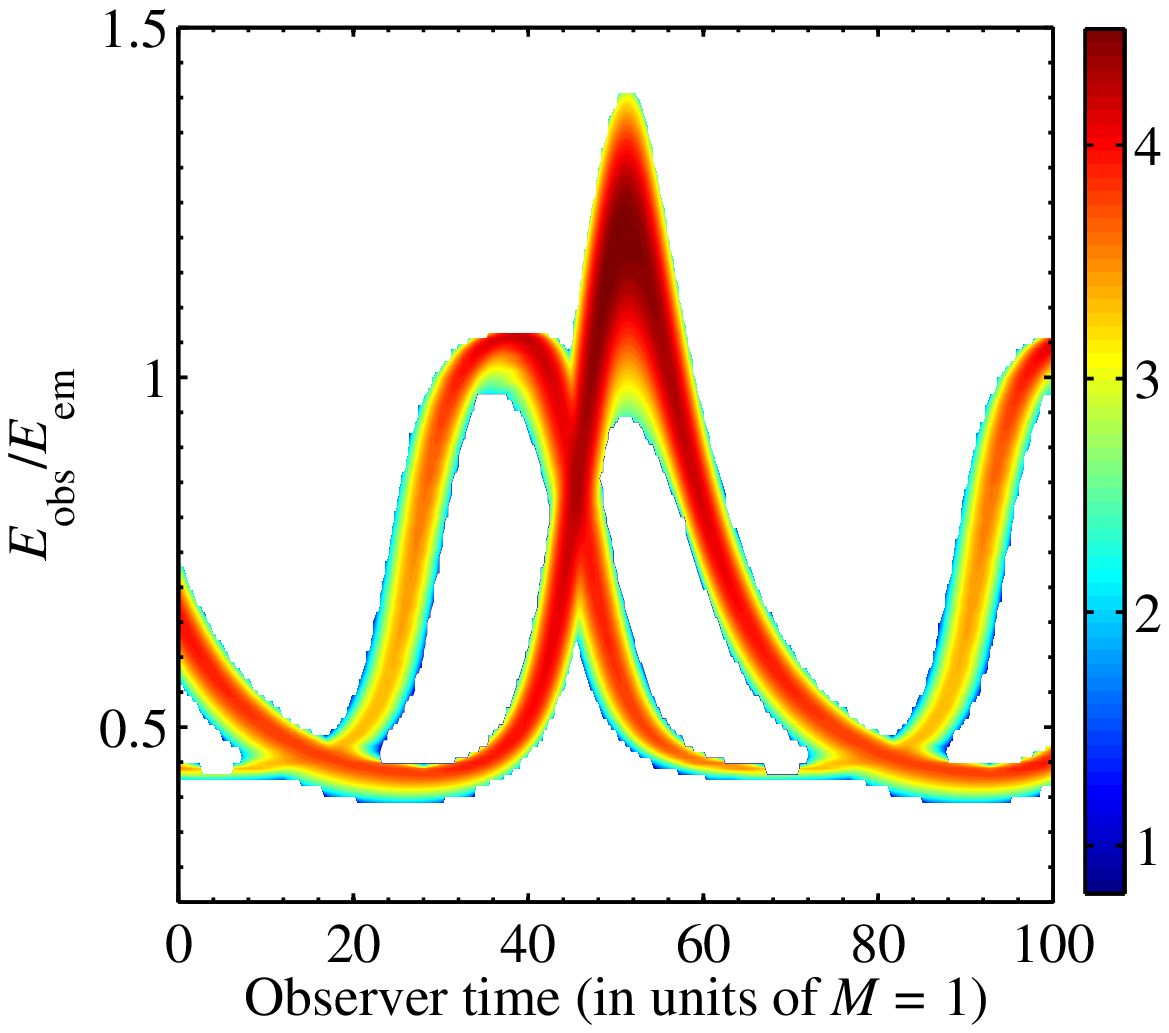}\\
\includegraphics[height=7cm]{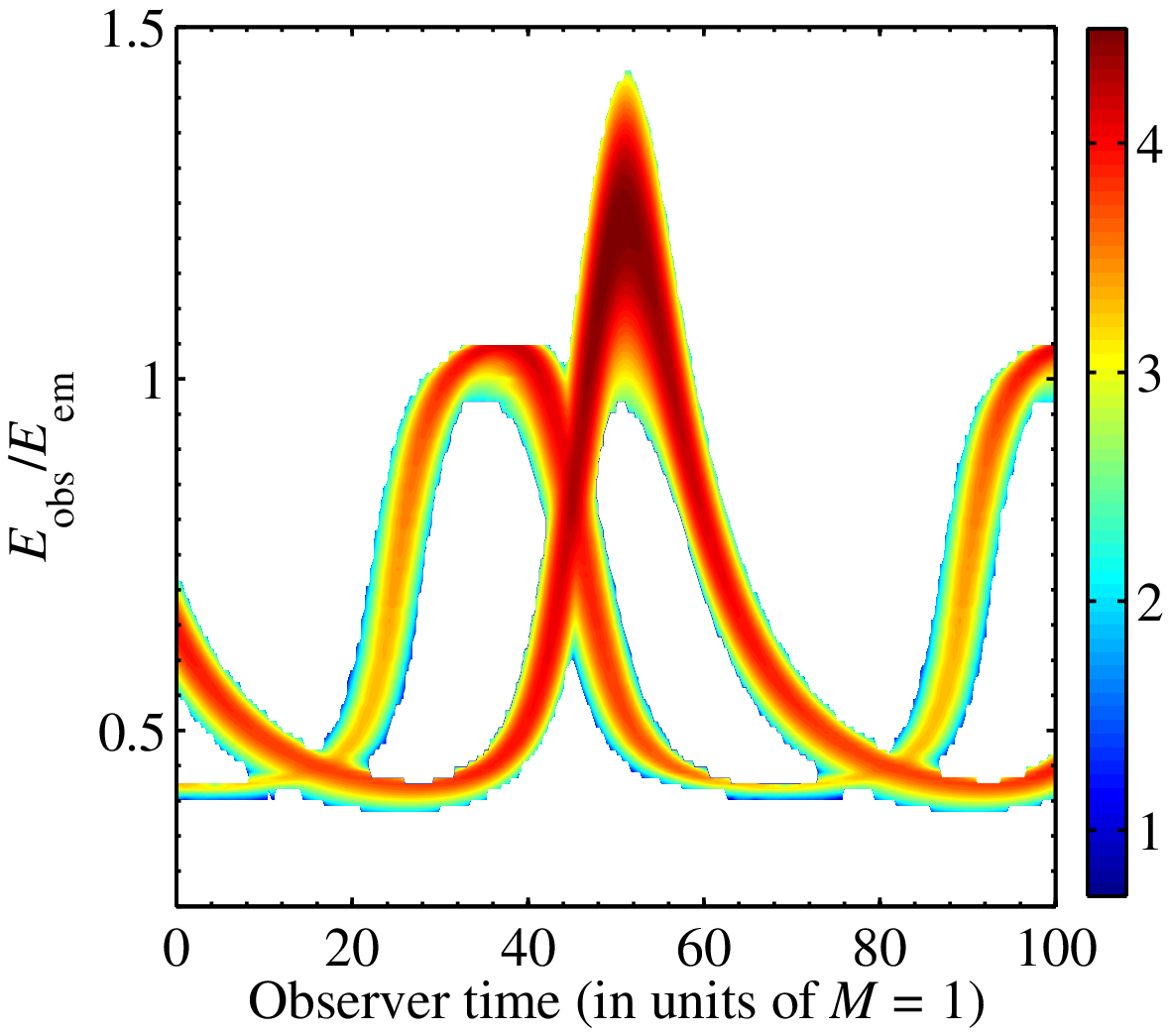}
\includegraphics[height=7cm]{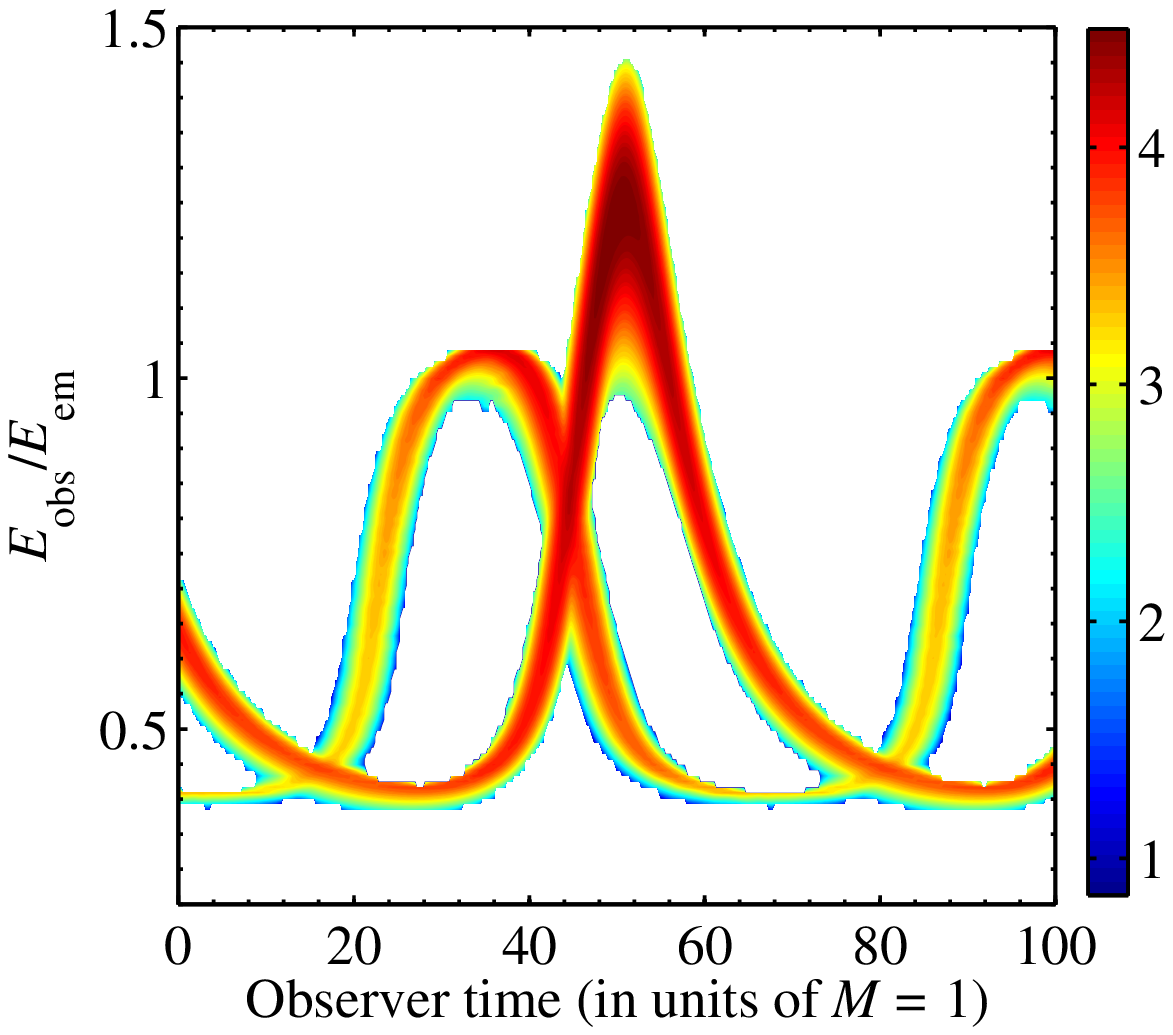}
\end{center}
\caption{Hot spot spectrograms for $a/M = 0.04$ and $\epsilon_3 = 8$ (top left 
panel), $a/M = 0.16$ and $\epsilon_3 = 4$ (top right panel), $a/M = 0.4$ and 
$\epsilon_3 = 8$ (bottom left panel), $a/M = 0.72$ and $\epsilon_3 = -3$ (bottom 
right panel). These spacetimes correspond to the ones marked by a green star in
the top left panel in Fig.~\ref{fig14} and to the one of the simulated light curve. 
See the text for more details.
\label{fig8b}}
\end{figure*}

\begin{figure}
\begin{center}
\includegraphics[height=7cm]{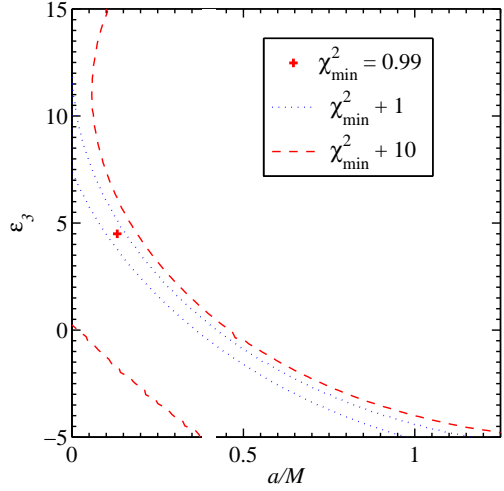}
\end{center}
\caption{As in the top left panel of Fig.~\ref{fig14} in the case of a simulated light curve 
in Johannsen-Psaltis spacetime with $\tilde{a}/M = 0.16$  and $\tilde{\epsilon}_3 = 4$. 
\label{fig15}}
\end{figure}

\begin{figure}
\begin{center}
\includegraphics[height=7cm]{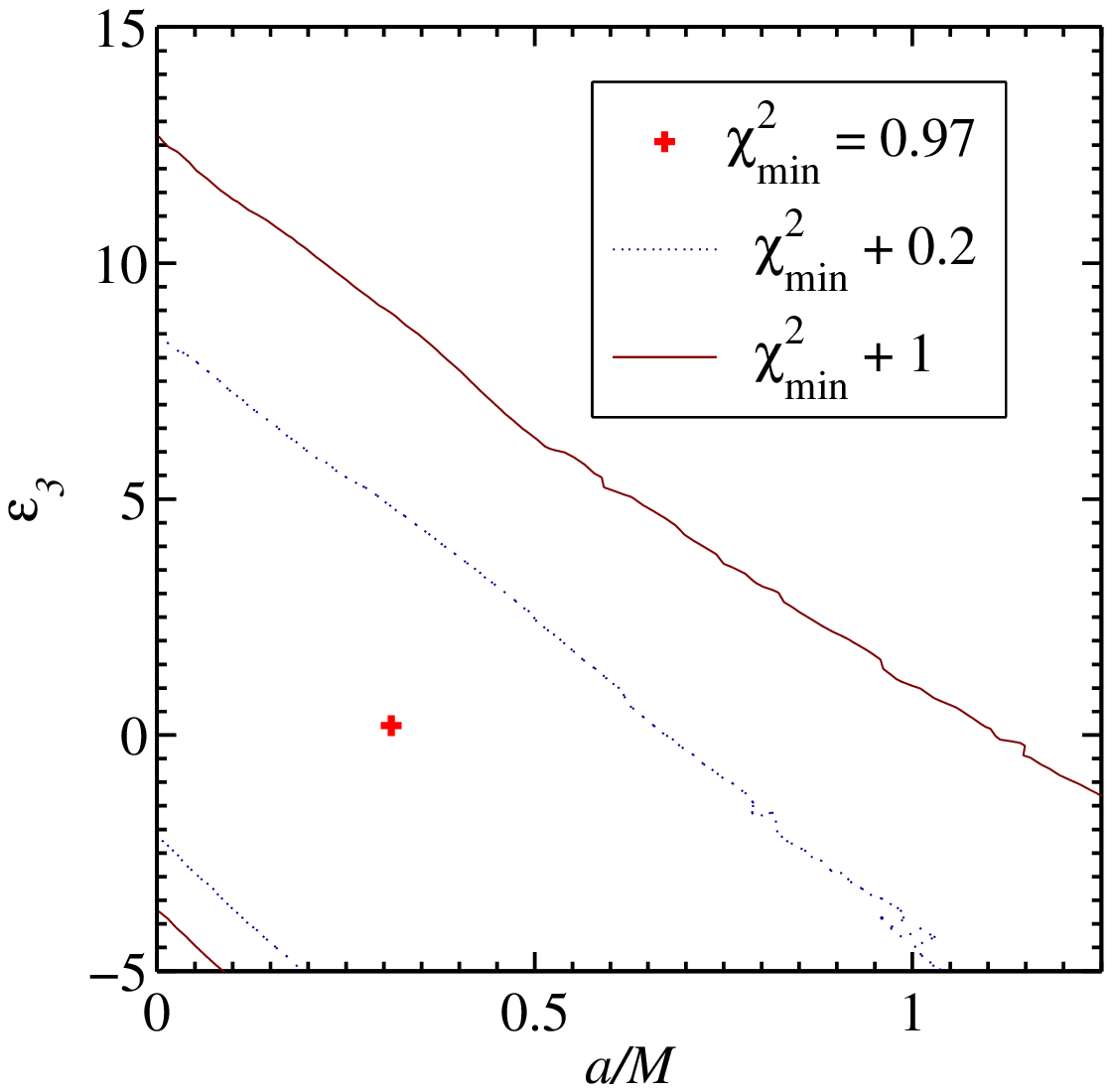}
\includegraphics[height=7cm]{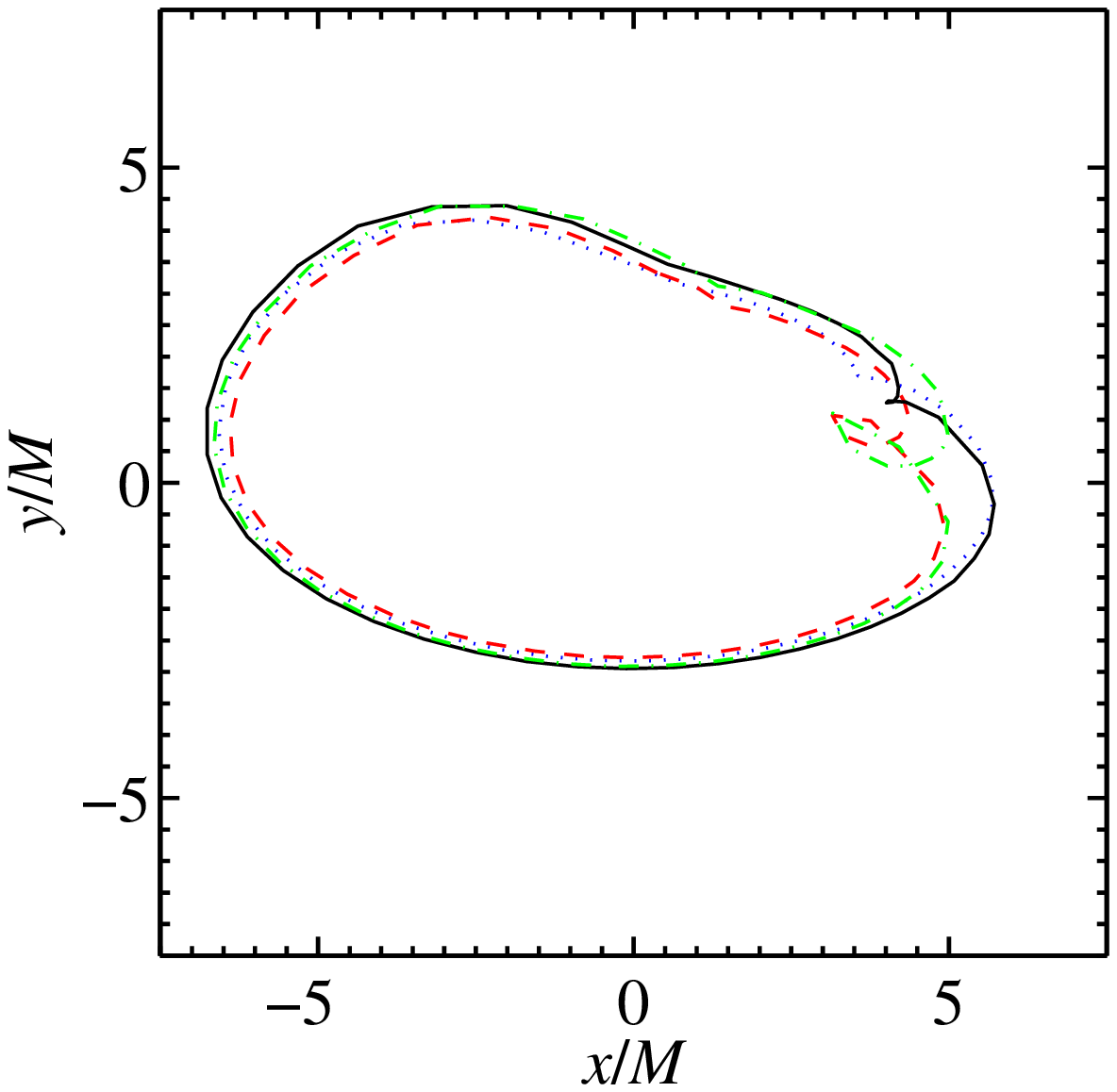}
\end{center}
\caption{Left panel: $\chi^2_{\rm red}$ from the comparison of the simulated total 
light curve of a hot spot orbiting a Kerr BH with $\tilde{a}/M = 0.4$ at the radius whose 
orbital frequency is the same as the one of the ISCO of a Schwarzschild BH and 
the theoretical light curves expected from hot spots orbiting Johannsen-Psaltis 
BHs at a radius with the same orbital frequency. The hot spot size is $R_{\rm spot} 
= 0.3 \, M$ and the observer's viewing angle is $i = 60^\circ$. Right panel: centroid
tracks of hot spots orbiting around BHs with $a/M=0.4$ and $\epsilon_3=0$ (black 
solid curve), $a/M=0.7$ and $\epsilon_3=0$ (green dashed-dotted curve), 
$a/M=0.4$ and $\epsilon_3=5$ (red dashed curve), $a/M=0.1$ and $\epsilon_3=5$ 
(blue dotted curve) at a radius with the same orbital frequency as the one of the 
ISCO of a Schwarzschild BH. See the text for more details. \label{fig16}}
\end{figure}

The case of hot spots orbiting non-Kerr BHs at radii with the same orbital frequency
is shown in Fig.~\ref{fig7}. As in the Kerr background in Fig.~\ref{fig5b}, light
curves with the same period have a slightly different shape due to the different
background metric. The left panel in Fig.~\ref{fig7} shows the total light curve of
hot spots orbiting around non-Kerr BHs with $a/M = 0.5$ for different values of the
deformation parameter $\epsilon_3$. The right panel compares instead the light
curve associated to hot spots orbiting non-Kerr BHs with $\epsilon_3 = 3$ and
different values of the spin $a/M$. In conclusion, it seems that the spin and possible
deviations from the Kerr solution affect the light curve of hot spots in a quite similar
way. In the next section, we will be more quantitative, and we will see how the
spin and the deformation parameter $\epsilon_3$ are correlated, as well as if
and how it is possible to get an independent estimate of them.

\vspace{2cm}

\section{Discussion}
\label{s-d}

The light curve of a hot spot is characterized by a time scale, set by the hot spot 
orbital period, and by a magnification, mainly determined by the orbital inclination 
with respect to the line of sight of the distant observer. The spin and possible 
deviations from the Kerr solution set the ISCO radius and therefore the ISCO 
frequency. The exact shape of the light curve also depends on the background 
metric, but the effects are smaller. In order to be more quantitative and figure out 
the information carried by the light curve, we can compare light curves produced 
in Kerr and non-Kerr spacetimes. We define the reduced $\chi^2$ as
\begin{align}
\chi^2_{\rm red}(a/M,\epsilon_3) =\frac{\chi^2}{n-2}
=\frac{1}{n-2}\sum^n_{i=1}\frac{[L_{i}^{\rm Sim}(\tilde{a}/M,\tilde{\epsilon}_3)
-L_{i}^{\rm Th}(a/M,\epsilon_3)]^2}{\sigma^2},
\end{align}
where the summation is performed over $n=50$ sampling times $t_i$. $L_{i}^{\rm Sim}$
is a simulated (i.e. noisy) normalized light curve calculated in the spacetime with
spin $\tilde{a}/M$ and deformation parameter $\tilde{\epsilon}_3$. Such a light 
curve is generated by assuming a Gaussian noise of 10\%. $L_{i}^{\rm Th}$ is
instead the theoretical prediction for a light curve in a spacetime with spin $a/M$
and deformation parameter $\epsilon_3$. 
The error $\sigma$ is assumed to be $4\%$ of the value of the peak of 
$L^{\rm Sim}$ and it does not depend on $L_{i}^{\rm Sim}$ because the amplitude
of the quasi-periodic modulation is much smaller than the mean flux of the flare.
Such a value roughly corresponds to the error of current observations with the ESO-VLT 
facility; see e.g. the figures in \citet{v2-trippe} or in \citet{v2-model4}. In the case of 
higher or lower values of $\sigma$, one has just to rescale $\chi^2_{\rm red}$, 
without altering its shape.

The mass of SgrA$^*$ is estimated to be around 4~million Solar masses, with an 
uncertainty of about 10\%. Even assuming the Kerr metric, both the spin and the
inclination angle are not really constrained to date, in the sense that different
authors give quite different results. Cosmological evolution arguments suggest
that, generally speaking, supermassive BHs are today not rotating rapidly~\citep{v2-spin}. 
However, the case of SgrA$^*$ is more controversial and actually many 
GRMHD simulations give high spin best fits, see e.g.~\citet{v2-drappeau}. The 
inclination of SgrA* cannot be straightforwardly derived from the magnification 
of the hot spot, as the observed data show a complicated signal, which is the 
superposition of a plausible sinusoidal component (that may be a hot spot) 
and a gaussian envelope. In what follows, we consider a specific case as an
example to illustrate the main purpose of the present work, which is how the
observation of a hot spot can be used to test the Kerr nature of SgrA$^*$,
without having in mind specific values of the BH parameters.

Let us first assume that SgrA$^*$ is a Kerr BH with $\tilde{a}/M = 0.4$ and that we
observe a hot spot orbiting at the ISCO radius. The simulated light curve is
shown in Fig.~\ref{fig13}. For the sake of simplicity, we fix the observer's inclination 
angle $i=60^\circ$ and the hot spot size $R_{\rm spot} = 0.3 \, M$. The reduced 
$\chi^2$ is reported in the top left panel of Fig.~\ref{fig14}, which shows
that the hot spot light curve can only select those spacetimes with the same ISCO
frequencies, while the differences due to the background metric are too small
to provide any useful information. In such a situation, one would like thus to figure
out if additional observations can break the correlation between spin and possible
deviations from the Kerr geometry. The VLTI instrument GRAVITY will be able 
to image hot spot orbiting around SgrA$^*$ with an angular resolution of about 
10~$\mu$as (which correspond to about $2M$) and a time resolution of 
about 1~minute. The three green stars in the top left panel of Fig.~\ref{fig14} 
correspond to the minima of the reduced $\chi^2$ for $\epsilon_3 = 8$,
4, -3. The centroid tracks in these spacetimes and the one in a Kerr background
corresponding to the simulated light curve are reported in the top right panel of
Fig.~\ref{fig14} [the centroid at any time is calculated following~\citet{v2-model4}]. 
The ISCO frequency of these backgrounds is essentially the same
and the four light curves are practically indistinguishable. The shape of the centroid
tracks is different. The origin of this difference is the contribution of the secondary 
image, whose peak phase shift with respect to the peak of the primary image 
changes a little bit with the background metric (see Fig.~\ref{fig8b}). 
Considering that the differences among these centroid tracks is not larger 
than $M$ and that we are assuming a quite simple hot spot model (the emission 
will not be really monochromatic and the hot spot will likely be stretched due to
shearing close to the BH), the GRAVITY instrument will be unable to distinguish 
these spacetimes.
The effect of the BH mass uncertainty is shown in the bottom panels of Fig.~\ref{fig14}.
If the BH mass is larger (smaller) than 10\%, the ISCO frequency found for a Kerr
BH with $\tilde{a}/M = 0.4$ corresponds instead to a Kerr BH with $\tilde{a}/M\approx 0.5$
(0.3). The bottom panels in Fig.~\ref{fig14} show the reduced $\chi^2$ for a simulated
light curve calculated in a Kerr background with $\tilde{a}/M = 0.3$ (left panel) and
0.5 (right panel).

A more attracting possibility to test the spacetime geometry around SgrA$^*$ is to
combine the hot spot information with the measurements of the BH mass and spin
that could be obtained from accurate radio observations of a radio pulsar in a 
compact orbit around SgrA$^*$. At present, no similar pulsar is known, but there
are a lot of efforts to find pulsars around the supermassive BH candidate at the
Center of our Galaxy and their discovery is probably only an issue of time. 
A pulsar with an orbital period of a few months would provide very accurate 
measurements of the BH mass and spin, as discussed in~\citet{v2-pulsar2}. The
key-point here is that the pulsar is in the weak gravitational field of SgrA$^*$,
where the mass is the monopole term of the gravitational field, the spin is the 
dipole term, and deviations from the Kerr solutions would appear at higher 
orders. The pulsar measurement therefore provides the actual spin, independently 
if it is orbiting around a Kerr or non-Kerr BH. The yellow vertical line in Fig.~\ref{fig14}
is the possible spin measurement $a/M = 0.40 \pm 0.01$ from a similar radio pulsar.
The combination of these measurement with the ISCO frequency of the hot spot
would result in a quite small allowed area; that is, possible deviations from the Kerr 
solutions can be strongly constrained. Here one may argue that the sole observation
of the radio pulsar may test the quadrupole moment of SgrA$^*$ and therefore
check if it is consistent with the predictions of the Kerr metric~\citep{v2-pulsar2}.
While that is definitively true, there are two significant advantages with the use of the
hot spot information. First, the pulsar may only test the quadrupole moment of
SgrA$^*$, while this object may have the same quadrupole moment as a Kerr BH 
and have deviations starting from higher order moments. The hot spot near the
ISCO would be sensitive to these higher order deviations. Second, the pulsar can test the quadrupole moment of SgrA$^*$ only in the case it is a very fast-rotating 
BH~\citep{v2-pulsar2}. The hot spot does not require it.

Let us now consider the possibility that SgrA$^*$ is a Johannsen-Psaltis with 
$\tilde{a}/M = 0.16$ and $\tilde{\epsilon}_3 = 4$. The reduced $\chi^2$ is shown in
Fig.~\ref{fig15}. As it could be expected, the hot spot light curve can only select
the spacetime with the same ISCO frequency, and the reduced $\chi^2$ in
Fig.~\ref{fig15} has exactly the same structure as the ones in Fig.~\ref{fig14}.

We note that our hot spot model relies on two important assumptions. First, the hot
spot is supposed to orbit at the ISCO radius. Assuming that the hot spot model
is correct, present data show that the timescale of the substructure of the flares
ranges from 13 to 30~minutes. That would suggest that the hot spot is not 
necessary at the ISCO radius, but the clumps of matter can form at different 
radii before being destroyed after a few orbits~\citep{v2-trippe}. We may argue
that one could consider the shortest timescale never observed as a proxy of
the ISCO period, but this is still a strong assumption. To be more fair, even the
shortest timescale would provide an upper bound to the ISCO period. One
can relax the assumption that the hot spot is at the ISCO and consider the more
general case with the hot spot radius as a free parameter, but in this case it is 
quite difficult to put constraints on the model. The second important assumption
is the monochromatic emission of the spot. That does not affect the hot spot 
frequency, but it plays a major
role in the calculation of the hot spot image and
therefore in the prediction of the centroid track
[it significantly changes the redshift factor, see e.g. Eq.~(6) in \citet{v2-model4}]. 
Here we have just considered
the simplest case and if one really wants to use the centroid track to constrain the
geometry around SgrA$^*$ a more detailed analysis with more realistic 
emission models would be necessary.

If we relax the assumption that the hot spot is orbiting at the ISCO radius, the 
estimate of $a/M$ and $\epsilon_3$ is more challenging. The reduced $\chi^2$ 
of this case is shown in the left panel of Fig.~\ref{fig16}. The simulated light curve
is calculated in a Kerr spacetime with $\tilde{a}/M = 0.4$. The hot spot orbital
frequency is the same as the one at the ISCO of a Schwarzschild BH. As in the
previous examples, we introduced a Gaussian noise of 10\%. The theoretical light
curves are computed for any spacetime at the radius with an orbital frequency
equal to the one of the simulated light curves. The final result is that the hot spot
cannot put any constraint
on the spin-deformation
parameters plane. In the end, it is true that one cannot really extract much more
information than the orbital frequency. 
In Fig.~\ref{fig16}, we also report the curve $\chi^2_{\rm min} + 0.2$ just to better
show the shape of the reduced $\chi^2$.
If we consider the centroid tracks of different
spacetimes, it turns out that the shape of the track is different. This is reported in the 
right panel of Fig.~\ref{fig16}, which shows the cases $a/M=0.4$ and $\epsilon_3=0$ 
(black solid curve), $a/M=0.7$ and $\epsilon_3=0$ (green dashed-dotted curve), 
$a/M=0.4$ and $\epsilon_3=5$ (red dashed curve), $a/M=0.1$ and $\epsilon_3=5$ 
(blue dotted curve). 
However, as in the previous case of a hot spot at the ISCO radius, 
GRAVITY will not be able to distinguish these metrics.

\vspace{1cm}

\section{Summary and conclusions}
\label{s-c}

SgrA$^*$, the supermassive BH candidate at the Center of our Galaxy, shows
X-ray and NIR flares of 1-3 hours at a rate of a few per day. Every flare has a
quasi-periodic substructure with a time scale of about 20~minutes. 
The hot spot model explains such a substructure with
the presence of a compact
emitting region orbiting near the ISCO of SgrA$^*$. This scenario can be
hopefully confirmed by the near future VLTI instrument GRAVITY, which has
the capability to image the hot spot and observe its motion. The possibility
of observing a bright compact source near the ISCO of SgrA$^*$ can be used
to probe the spacetime geometry around this object and verify if it is a Kerr
BH, as expected in general relativity.

In the present paper, we have extended previous results found in the Kerr 
background to the case of non-Kerr metrics, to figure out how the observation 
of a hot spot near the ISCO radius can test the Kerr nature of SgrA$^*$. As in 
the case of other techniques, the main problem is the strong correlation between 
effects due to the spin and the ones coming from possible deviations from 
the Kerr solution. Since the BH mass is measured by dynamical methods, by 
studying the orbital motion of individual stars at relatively large distances from 
the compact object and in the framework of Newtonian gravity, the spin is the 
only background parameter if we assume that SgrA$^*$ is a Kerr BH. Any 
observable quantity determined by the geometry of the spacetime can be used 
to infer the value of the spin. If we want to test the Kerr metric of SgrA$^*$, in 
addition to the spin we have a deformation parameter, which is used to quantify 
possible deviations from the Kerr solution. Now the observable quantity 
depends on two parameters and it is not easy to get an independent estimate 
of both.

If we assume that the quasi-periodic substructure with the shortest timescale
is produced by hot spots at the ISCO of SgrA$^*$, the hot spot light curve
provides the ISCO frequency. In the case of a non-Kerr background with one 
deformation parameter, one finds an allowed region on the spin-deformation 
parameter plane. Light curves in different metric are not exactly the same, 
but the difference is so small that it cannot be identified with current and near 
future facilities, so eventually only the period of the hot spot matters. For this
reason, if we want to test the Kerr nature of SgrA$^*$ we need additional
measurements. GRAVITY will be able to observe the orbital motion of a hot spot
orbiting near the ISCO of the supermassive BH candidate at the Center of 
the Galaxy with an angular resolution of order 10~$\mu$as and a time resolution 
of about 1~minute. If the hot spot model is correct, the centroid track depends on 
the background metric and BH spacetimes with the same ISCO frequency have 
centroid tracks with different shape, which means that the simultaneous 
measurement of the orbital frequency and of the image of the hot spot can 
potentially break 
the degeneracy between spin and possible deviations from the Kerr geometry.
However, that is out of reach for GRAVITY.
A more appealing possibility is represented by the discovery
of a radio pulsar in a compact orbit (orbital period of a few months) around 
SgrA$^*$. In such a case, accurate radio observations would be able to get
precise measurements of the BH mass and spin in the weak field regime,
independently of possible deviations from the Kerr solution. The combination 
of these measurements with the hot spot constraint would provide a stringent 
test of the Kerr BH hypothesis.


\begin{acknowledgments}
We thank Rohta Takahashi for useful discussions and suggestions.
This work was supported by the NSFC grant No.~11305038, the 
Shanghai Municipal Education Commission grant for Innovative 
Programs No.~14ZZ001, the Thousand Young Talents Program, 
and Fudan University.
\end{acknowledgments}


\end{document}